\documentclass[aps,pra,reprint,nofootinbib,superscriptaddress,twocolumn,showpacs,showkeys,longbibliography,amsmath,amssymb]{revtex4-2}
\usepackage{graphicx}% Include figure files
\usepackage{dcolumn}% Align table columns on decimal point
\usepackage{bm}% bold math
\usepackage{braket}
\usepackage{subfigure}
\usepackage[colorlinks,bookmarks=false,citecolor=blue,linkcolor=red,urlcolor=blue]{hyperref}
\usepackage[english]{babel}
\usepackage{changes}
\usepackage{multirow}
\usepackage{booktabs}
\usepackage{float}
\usepackage{array} 
\begin{document}
\preprint{APS/123-QED}	
	
\title{Self-consistent effective field theory to nonuniversal Lee-Huang-Yang term in quantum droplets}
\author{Yi Zhang}
\affiliation{Department of Physics, Zhejiang Normal University, Jinhua 321004, People's Republic of China}
\author{Xiaoran Ye}
\affiliation{Department of Physics, Zhejiang Normal University, Jinhua 321004, People's Republic of China}
\author{Ziheng Zhou}
\affiliation{Department of Physics, Zhejiang Normal University, Jinhua 321004, People's Republic of China}
\author{Zhaoxin Liang}\email[Corresponding author:~] {zhxliang@zjnu.edu.cn}
\affiliation{Department of Physics, Zhejiang Normal University, Jinhua 321004, People's Republic of China}

\date{\today}	

\begin{abstract}

Quantum droplets (QDs) in weakly interacting ultracold quantum gases are typically characterized by mean-field theories incorporating Lee-Huang-Yang (LHY) quantum fluctuations under simplified zero-range interaction assumptions. However, bridging these models to broader physical regimes like superfluid helium requires precise understanding of short-range interatomic interactions. Here, we investigate how finite-range interactions—next-order corrections to zero-range potentials—significantly alter QDs mechanics. Using a consistent effective theory, we derive an analytical equation of state (EOS) for three-dimensional bosonic mixtures under finite-range interactions at zero temperature. Leveraging the Hubbard-Stratonovich transformation, we demonstrate that interspecies attraction facilitates bosonic pairing across components characterized by the non-perturbative parameter of $\Delta$,  leading to nonuniversal LHY terms that encode short-range interaction details while recovering previous universal QDs EOS in the zero-range limit. Extending superfluid hydrodynamic equations for two-component systems, we predict fractional frequency shifts in breathing modes induced by these nonuniversal terms. Experimental observation of these shifts would reveal critical insights into QDs dynamics and interatomic potential characteristics.
\end{abstract}
\maketitle

\section{INTRODUCTION\label{1}}

Quantum droplets (QDs)~\cite{Bulgac2002,Ferrier2016,Science2018,Schmitt2016,Hu2020}—self-bound clusters of tens to hundreds of thousands of atoms—emerge from a balance between attractive mean-field interactions and repulsive quantum fluctuations~\cite{Nagaosa1999,Pitaevskii2016,Peskin1995}. This stabilization mechanism~\cite{Petrov2015}, analogous to classical superfluid helium~\cite{McMillan1965,Harms1998,Dalfovo2001,Francesco2017,Pang2020}, has established a novel platform for quantum many-body physics~\cite{Bloch2008,Boudjemaa2018,Abbas2023}. Recent theoretical advances~\cite{Petrov2015,Petrov2016} demonstrate that three-dimensional (3D) bosonic mixtures with intraspecies repulsions (\(a_{11}\,,a_{22}>0\)) and interspecies attractions (\(a_{12}<0\)) form self-bound QDs through modified Bogoliubov theory, where quantum fluctuations counteract collapse~\cite{HeLi2023}. However, existing zero-range interaction models neglect critical short-range effects that dominate real-world systems. Nonuniversal finite-range corrections~\cite{Dalfovo1999,Salasnich2017,Cappe2017,Cappellaro2017,Tononi2018,Yin2020,Yu2024,Ye2024,Zhang2024,Emerson2025}—essential for stabilizing classical droplets~\cite{Schmitt2016,Petrov2018,Chiquillo2019,Zhang2025,Flachi2025,Englezos2025}—modify the Lee-Huang-Yang (LHY) energy functional by introducing density-dependent terms that alter equilibrium properties. These corrections are particularly significant in ultracold atomic gases, where precise scattering length measurements and anisotropy effects are crucial for interpreting experimental observations~\cite{Ferioli2020,Emerson2025,Banerjee2025,Cavicchioli2025,Xiaoran2025}.

Within the theoretical framework of quantum droplets, Petrov's seminal work~\cite{Petrov2015} provides a foundational description by integrating mean-field theory with LHY quantum fluctuations~\cite{Lee1957,Yang1957}. However, this approach presents a challenge near collapse phases where interspecies interactions satisfy $\delta a = a_{12} + \sqrt{a_{11}a_{22}} < 0$~\cite{Science2018}. Specifically, the softening of one Bogoliubov spectrum under collapse conditions leads to complex eigenvalues, which introduces challenges in maintaining quantum phase stability. Recent theoretical advances by Hui Hu's group~\cite{Hu2020,Hu2020b,Hu2020d,Hu2025} have revisited Petrov's seminal work~\cite{Petrov2015} through a Hubbard-Stratonovich (HS) transformation approach. This formalism reveals that bosonic pairing dynamics, analogous to Cooper pairing in Bardeen-Cooper-Schrieffer (BCS) superconductors~\cite{Bardeen1957,Hu2006,He2015,Shen2025}, can emerge in two-component Bose-Bose mixtures under attractive interspecies interactions. Crucially, the formalism developed by Hui Hu's group~\cite{Hu2020,Hu2020b,Hu2020d,Hu2025} yields an analytically improved LHY correction that aligns with diffusion Monte Carlo (DMC) simulations~\cite{Petrov2016,Cikojevi2019,Cikojevi2020,Cikojevi2020b,Parisi2019}, resolving earlier discrepancies observed in collapse-phase predictions. Nevertheless, practical implementation faces challenges when bridging theory to experimental systems. While the current formalism assumes contact interactions~\cite{Hu2020,Hu2020b}, real atomic systems exhibit finite-range interaction profiles~\cite{Lorenzi2023} with characteristic short-range details that require further theoretical incorporation.

In this work, we are motivated to present a unified framework employing bosonic pairing theory to investigate ground-state properties of 3D bosonic mixtures under finite-range interactions at zero temperature. Within the HS transformation framework, our approach relies on interspecies attraction inducing bosonic pairing across components, parameterized by the non-perturbative parameter $\Delta \propto 1/\delta a$. This formalism establishes a non-perturbative theory that conceptually extends Petrov's seminal treatment~\cite{Petrov2015} where $\delta a$ is treated as a small parameter. This formalism systematically derives an analytical equation of state (EOS) incorporating both mean-field and quantum fluctuation effects, enabling precise characterization of nonuniversal energy per particle and equilibrium density in QDs. By tuning interatomic scattering lengths through magnetic Feshbach resonances, we demonstrate the emergence of a self-bound QDs regime where interspecies attraction counterbalances 3D quantum pressure. Our results reconcile Hui Hu's universal QDs EOS~\cite{Hu2020,Hu2020b,Hu2025} with LHY quantum fluctuation corrections. Extending this, we derive a generalized hydrodynamic framework for binary mixtures by incorporating LHY-induced density fluctuations into superfluid equations of motion. Deriving these refined dynamics predicts fractional frequency shifts in breathing-mode oscillations—a direct signature of the mixture’s nonuniversal EOS. Accessible in current cold-atom experiments, these shifts probe QDs self-binding. Observing them would mark a paradigm shift in QDs studies, enabling direct tests of quantum-fluctuation-stabilized mechanisms and new explorations in beyond-mean-field quantum hydrodynamics.

The paper is organized as follows. Section~\ref{2} constructs the thermodynamic potential for 3D two-component bosonic mixtures under finite-range interactions within a self-consistent field theory framework. Building on this, Sec.~\ref{3} derives explicit expressions for energy per particle and equilibrium density through analytical solutions of the governing equations. Section~\ref{4} reveals fractional breathing-mode frequency shifts induced by self-consistent LHY corrections to the nonuniversal chemical potential. Finally, Sec.~\ref{5} summarizes key findings and outlines experimental conditions for observing these QDs phenomena.

\section{FUNCTIONAL PATH INTEGRAL WITH BOSONIC PAIRING THEORY\label{2}}
\subsection{Partition function of the system}
\label{IIA}
In this work, we focus on a two-component 3D weakly interacting bosonic mixtures~\cite{Emerson2025,Chiquillo2018} with finite-range interactions~\cite{Cappe2017,Salasnich2017}. To analyze this system, we adopt the path-integral formalism~\cite{Nagaosa1999,Pitaevskii2016,Peskin1995} for which the Euclidean partition function of the model reads~\cite{Emerson2025,Chiquillo2018}
\begin{eqnarray}
	\label{partition function}
	\mathcal{Z} & = & \int\mathcal{D}\left[\Phi,\Phi^{*}\right]\exp\left\{-\frac{S\left[\Phi,\Phi^{*}\right]}{\hbar}\right\},
\end{eqnarray}
with the Euclidean action functional defined as $S[\Phi,\Phi^*] = \int_0^{\beta\hbar} d\tau \int_{\mathcal{V}} d\mathbf{r}\,\mathcal{L}[\Phi,\Phi^*]$, where $\beta = 1/(k_\text{B}T)$ (inverse thermal energy scale) and $\mathcal{V}$ is the system volume. The Lagrangian density $\mathcal{L}$, governing dynamics, reads
\begin{eqnarray}
	\label{Lagrangian density}	
\mathcal{L}&=&\sum_{i=1,2}\Bigg\{\phi_{i}^{*}\left[\hbar\partial_{\tau}-\frac{\hbar^{2}\nabla^{2}}{2m}-\mu_{i}\right]\phi_{i}\nonumber\\
&+&\frac{g_{ii}^{\left(0\right)}}{2}\left|\phi_{i}\right|^{4}-\frac{g_{ii}^{\left(2\right)}}{2}\left|\phi_{i}\right|^{2}\nabla^{2}\left|\phi_{i}\right|^{2}\Biggr\}
\nonumber\\
&+&g_{12}^{\left(0\right)}\left|\phi_{1}\right|^{2}\left|\phi_{2}\right|^{2}-g_{12}^{\left(2\right)}\left|\phi_{1}\right|^{2}\nabla^{2}\left|\phi_{2}\right|^{2}.
\label{Ldensity}
\end{eqnarray}

In Equation~(\ref{Lagrangian density}), the two-component complex field $\Phi(\mathbf{r},\tau)=[\phi_1,\phi_2]^\text{T}$ describes bosonic degrees of freedom varying in the space of $\mathbf{r}$ and the imaginary time of $\tau$, where $\phi_{1,2}$ are the mixture components and $\mu_i$ is the chemical potential of species $i$.  
$g_{ij}^{(0)}$ are intra- ($i=j$) and inter-species ($i\neq j$) zero-range couplings, regularized via s-wave scattering lengths. Crucially, $g_{ij}^{(2)}$ in Eq.~(\ref{Lagrangian density}) encode finite-range effects, arising from an s-wave pseudopotential expansion with a finite-range length scale that extends the zero-range model to include short-range correlations.  
In 3D, these couplings are given by~\cite{Lorenzi2023}  
\begin{equation}
	g_{ij}^{(0)} = \frac{4\pi\hbar^2 a_{\text{s}_{ij}}}{m}, \quad 
	g_{ij}^{(2)} = \frac{2\pi\hbar^2 a_{\text{s}_{ij}}^2 r_{\text{eff}}^{\text{s}_{ij}}}{m}.\nonumber
\end{equation}  
Here, $a_{\text{s}_{ij}}$
represents the s-wave scattering length, and $r_{\text{eff}}^{\text{s}_{ij}}$ is the finite-range length of the pseudopotential. The effective range $r_{\text{eff}}^{\text{s}_{ij}}$, defined via the low-energy expansion of the s-wave phase shift, corresponds to the scattering effective range (distinct from the physical range of the interatomic potential). For bosonic mixtures, $r_{\text{eff}}^{\text{s}_{ij}}$ can take positive or negative values, depending on the interplay between intraspecies repulsion and interspecies attraction.

For context, Bogoliubov theory establishes the zero-temperature nonuniversal energy per particle of the Lagrangian density functional~\cite{Emerson2025}, where the thermodynamic potential is expressed as $\Omega[\phi_{i{\text{c}}},\mu]/\mathcal{V}$~\cite{Petrov2015} with $\phi_{i{\text{c}}}$ the condensate field. In contrast, this work employs bosonic pairing theory~\cite{Hu2020,Hu2020b,Hu2025} to derive a self-consistent nonuniversal EOS for QDs, yielding fully positive excitation spectra under finite-range interactions. A key step employs the HS transformation to compute the thermodynamic potential $\Omega[\phi_{i{\text{c}}},\Delta,\mu]/\mathcal{V}$. At the saddle point, $\delta\Omega/\delta\Delta = 0$ decouples interspecies interactions in the Lagrangian's pairing terms via the field $\Delta$. Combined with Refs.~\cite{Hu2020,Emerson2025}, our results consistently describe the nonuniversal energy per particle for 3D bosonic mixtures, incorporating finite-range effects of the interatomic potential.

Our subsequent goal in Sec.~\ref{2} is to systematically derive the consistent thermodynamic potential  
$\Omega\left[\phi_{i{\text{c}}},\Delta,\mu\right]/\mathcal{V}$
governed by the Lagrangian density functional~(\ref{Lagrangian density}), within the framework of self-consistent field theory~\cite{Peskin1995}. This is achieved via the HS transformation, which enables a beyond mean-field decoupling of the interaction terms while preserving the quantum statistical correlations. The resultant thermodynamic potential is then employed to compute the nonuniversal EOS~\cite{Cappellaro2017,Chiquillo2018,Emerson2025,Tononi2018,Tononi2018,Zhang2024,Ye2024,Salasnich2017} of the bosonic mixtures, explicitly incorporating intra- and interspecies interactions through the saddle-point approximation.  

\subsection{Hubbard–Stratonovich transformation}
\label{IIB}
In Sec.~\ref{IIA}, we have given the Lagrangian density in Eq.~(\ref{Lagrangian density}), incorporating finite-range effects via $g_{ij}^{(2)}$ for the bosonic mixture. Whereas previous works~\cite{Emerson2025,Ye2026} have derived the zero temperature nonuniversal per particle energy via Bogoliubov theory; Sec.~\ref{IIB} develops a self-consistent field theory~\cite{Hu2020,Hu2020b,Hu2025} within the framework of HS transformation to obtain the nonuniversal LHY potential $\Omega_{\text{LHY}}$. Using the HS transformation, we decouple interspecies interactions via pairing field $\Delta(\mathbf{r},\tau)$, while  
derive saddle-point equations for condensate $\phi_{i{\text{c}}}$ and pairing field $\Delta_0(\mathbf{r},\tau)$.

The purpose and emphasis of this work is to establishe a unified description of QDs stability by integrating mean-field physics with beyond-mean-field quantum fluctuations arising from finite-range interactions.  Central to this approach is the HS transformation, which employs Gaussian integration to linearize quartic interaction terms in the Lagrangian density (\ref{Lagrangian density}). This mathematical procedure introduces auxiliary fields $\Delta(\mathbf{r},\tau)$ that effectively decouple interspecies interactions, thereby reformulating complex many-body problems into tractable quadratic forms coupled with auxiliary fields.

%The resultant framework will provide a unified description of QDs stability, capturing both the mean-field physics and beyond-mean-field quantum fluctuations arising from finite-range interactions.  
%The key idea of HS transformation is that, 
%by performing Gaussian integrals to transform exponential quadratic terms in Lagrangian density (\ref{Lagrangian density}) into linear terms, auxiliary fields $\Delta\left(\mathbf{r},\tau\right)$ are introduced to decouple interspecies interactions. This method leverages the mathematical properties of Gaussian integration to linearize quartic interactions, enabling the reformulation of complex many-body problems into tractable quadratic forms coupled with auxiliary fields.

First, the application of the HS framework in QDs originates from the necessity to address interspecies pairing interactions. Specifically, the attractive interspecies interaction in the Lagrangian density functional~(\ref{Lagrangian density}) motivates the HS transformation, which systematically decouples the dominant pairing mechanism encoded in the last two interaction terms. This is achieved through the introduction of a pairing field $\Delta(\mathbf{r},\tau)$ in the Lagrangian density functional~(\ref{Lagrangian density}), 
\begin{widetext}
\begin{equation}	
	\label{Lagrangian density with pairing field}
	e^{-\int_{0}^{\beta\hbar}d\tau\int_{\mathcal{V}} d\mathbf{r}\phi_{1}^{*}\phi_{2}^{*}\left(g_{12}^{\left(0\right)}-g_{12}^{\left(2\right)}\nabla^{2}\right)\phi_{2}\phi_{1}}
	=\int\mathcal{D}\left[\Delta\left(\mathbf{r},\tau\right)\right]\exp\left\{ \int_{0}^{\beta\hbar}d\tau\int_{\mathcal{V}} d\mathbf{r}\left[\frac{\left|\Delta\right|^{2}}{g_{12}^{\left(0\right)}-g_{12}^{\left(2\right)}\nabla^{2}}+\left(\Delta^{*}\phi_{2}\phi_{1}+\phi_{1}^{*}\phi_{2}^{*}\Delta\right)\right]\right\},
\end{equation}
\end{widetext}
effectively mapping the original fermionic problem onto an interacting bosonic field theory~\cite{Bardeen1957}.  
Replacing $e^{-\int_{0}^{\beta\hbar}d\tau\int_{\mathcal{V}} d\mathbf{r}\phi_{1}^{*}\phi_{2}^{*}\left(g_{12}^{\left(0\right)}-g_{12}^{\left(2\right)}\nabla^{2}\right)\phi_{2}\phi_{1}}$ in Eq.~(\ref{Ldensity}) with Eq.~(\ref{Lagrangian density with pairing field}), one can obtain the action $S=\int_{0}^{\beta\hbar}d\tau\int_{\mathcal{V}} d\mathbf{r}\mathcal{L}$ in the form of
\begin{eqnarray}
	\label{action with pairing field}
S&=&\int  \Bigg\{\Big[-\frac{\left|\Delta\right|^{2}}{g_{12}^{\left(0\right)}-g_{12}^{\left(2\right)}\nabla^{2}}-\left(\Delta^{*}\phi_{2}\phi_{1}+\phi_{1}^{*}\phi_{2}^{*}\Delta\right)\Big]\nonumber\\
&+&\sum_{i=1,2}\Big[\phi_{i}^{*}\left(\hbar\partial_{\tau}-\frac{\hbar^{2}\nabla^{2}}{2m}-\mu_{i}\right)\phi_{i}\nonumber\\
&+&\frac{g_{ii}^{\left(0\right)}}{2}\left|\phi_{i}\right|^{4}-\frac{g_{ii}^{\left(2\right)}}{2}\left|\phi_{i}\right|^{2}\nabla^{2}\left|\phi_{i}\right|^{2}\Big]\Bigg\}.
\end{eqnarray}

Next, for the pairing field $\Delta$, a uniform saddle-point solution $\Delta = \Delta_0 > 0$ is sufficient in our approximation. At this truncation level, we take the $i$-th component bosons to condense into zero momentum, decomposing $\phi_i(\mathbf{r},\tau) = \phi_{i{\text{c}}} + \delta\phi_i(\mathbf{r},\tau)$, where $\phi_{i{\text{c}}}$ is the condensate wavefunction~\cite{Lieb1963} and $\delta\phi_i$ describes fluctuations~\cite{Nagaosa1999,Pitaevskii2016} around the mean-field ground-state. For intraspecies interactions, we keep contributions up to one-loop level. These approximations yield the Lagrangian density in Eq.~(\ref{Lagrangian density}) separating into a mean-field part and fluctuation corrections:  
$\mathcal{L} = \mathcal{L}_0 + \mathcal{L}_2$ with
\begin{equation}
	\label{L0}
	\mathcal{L}_{0}=\sum_{i=1,2}\Big(-\mu_{i}\phi_{i{\text{c}}}^{2}+\frac{g_{ii}^{\left(0\right)}}{2}\phi_{i{\text{c}}}^{4}
	-\frac{\Delta^{2}}{g_{12}^{\left(0\right)}-g_{12}^{\left(2\right)}\nabla^{2}}-2\Delta\phi_{1{\text{c}}}\phi_{2\text{c}}\Big),
\end{equation}
and
\begin{eqnarray}
	\label{L2}
	\!\!\!\!\mathcal{L}_{2}&=&\sum_{i=1,2}\Big\{\delta\phi_{i}^{*}[\hbar\partial_{\tau}\!-\!\frac{\hbar^{2}\nabla^{2}}{2m}\!-\!\mu_{i}\!+\!\phi_{i{\text{c}}}^{2}(2g_{ii}^{\left(0\right)}-g_{ii}^{\left(2\right)}\nabla^{2})]\delta\phi_{i}\nonumber\\
	\!\!\!&+&\!\!\frac{\phi_{i{\text{c}}}^{2}}{2}[\delta\phi_{i}^{*}(g_{ii}^{\left(0\right)}\!-\!g_{ii}^{\left(2\right)}\nabla^{2})\delta\phi_{i}^{*}\!+\!\delta\phi_{i}(g_{ii}^{\left(0\right)}\!-\!g_{ii}^{\left(2\right)}\nabla^{2})\delta\phi_{i}]\Big\}\nonumber\\
	&-&\Delta\left(\delta\phi_{2}\delta\phi_{1}+\delta\phi_{1}^{*}\delta\phi_{2}^{*}\right).
\end{eqnarray}

At the leading order, the thermodynamic potential $\Omega_{0}=\mathcal{L}_{0}\mathcal{V}$ from condensates takes the form
\begin{equation}
	\label{omega0}
	\frac{\Omega_{0}}{\mathcal{V}}=-\frac{\Delta^{2}}{g_{12}^{\left(0\right)}-g_{12}^{\left(2\right)}\nabla^{2}}-2\Delta\phi_{1{\text{c}}}\phi_{2\text{c}}+\sum_{i=1,2}-\mu_{i}\phi_{i{\text{c}}}^{2}+\frac{g_{ii}^{\left(0\right)}}{2}\phi_{i{\text{c}}}^{4}.
\end{equation}
By defining $C_{i}=g_{ii}^{\left(0\right)}\phi_{i{\text{c}}}^{2}$ and minimizing $\Omega_{0}$ with respect to $\phi_{i{\text{c}}}$, i.e., $\delta\Omega_{0}/\delta \phi_{i{\text{c}}}=0$, we obtain $C_{i}=\mu_{i}+\Delta\left(\phi_{j\text{c}}/\phi_{i{\text{c}}}\right)$, $i\,,j=\left(1,2\right)$ and $\Omega_{0}/\mathcal{V}=-\Delta^{2}/(g_{12}^{\left(0\right)})-C_{1}^{2}/(2g_{11}^{\left(0\right)})-C_{2}^{2}/2(g_{22}^{\left(0\right)})$. 

The next-to-leading order contribution to the thermodynamic potential arises from Gaussian fluctuations around the condensate states, encoded in the bilinear Lagrangian density (\ref{L2}). To diagonalize the quadratic terms involving $\delta\phi_i$ and $\delta\phi_i^*$ in Eq.~(\ref{L2}), we first perform Fourier transformations on the fields ($\delta\phi_i$\,,  $\delta\phi_i^*$) to Matsubara frequencies ($\partial_\tau \to -i\omega_n$) and real-space momenta ($\nabla^2 \to -\mathbf{k}^2$), then work in Nambu space. This procedure reduces the action functional $S_{2}$ to a diagonal quadratic form, 
\begin{equation}
S_{2}\left[\delta \phi_{1}^{*},\delta \phi_{1},\delta \phi_{2}^{*},\delta \phi_{2}\right]=-\frac{\hbar}{2}\sum_{\mathbf{k},n}\mathbf{\Phi}_{\mathbf{k}n}\mathcal{D}^{-1}_{\mathbf{k}n}\mathbf{\Phi}_{\mathbf{k}n}^{\dagger},\label{S2}
\end{equation}
where $\hbar \mathbf{k}$ denotes the momentum and $n$ labels the Matsubara frequencies with $\omega_{n}=2\pi n/ \hbar\beta$. The vector in Eq.~(\ref{S2})
\begin{equation}
\mathbf{\Phi}_{\mathbf{k}n}=\left(\delta\phi_{1,\mathbf{k}n}^{*}\,,\delta\phi_{1,-\mathbf{k}n}\,,\delta\phi_{2,\mathbf{k}n}^{*}\,,\delta\phi_{2,-\mathbf{k}n}\right)\nonumber
\end{equation}
resides in the appropriate Nambu space, and the inverse Green's function of bosons in Eq.~(\ref{S2}), $\mathcal{D}^{-1}_{\mathbf{k}n}$, can be written as 
\begin{widetext}
	\begin{eqnarray}
	\mathcal{D}^{-1}\left(\mathbf{ k},\omega_{n}\right)&=\left[\begin{array}{cccc}
		\hbar\omega_{n}-B_{1\mathbf{k}} & -\left(C_{1}+D_{1}\mathbf {k}^{2}\right) & 0 & \Delta\\
		-\left(C_{1}+D_{1}\mathbf {k}^{2}\right) & -\hbar\omega_{n}-B_{1\mathbf{k}} & \Delta & 0\\
		0 & \Delta & \hbar\omega_{n}-B_{2\mathbf{k}} & -\left(C_{2}+D_{2}\mathbf{ k}^{2}\right)\\
		\Delta & 0 & -\left(C_{2}+D_{2}\mathbf{ k}^{2}\right) & -\hbar\omega_{n}-B_{2\mathbf{k}}
	\end{array}\right],
	\end{eqnarray}
\end{widetext}
where we have defined
$B_{i{\bf k}}=\varepsilon_{\mathbf{k}}-\mu_{i}+2C_{i}+D_{i}\text{{\bf k}}^{2}$ and $D_{i}=g_{ii}^{\left(2\right)}\phi_{i{\text{c}}}^{2}$, with $\varepsilon_{\mathbf{k}}$ being $\hbar^{2}\mathbf{k}^{2}/2m$.

By solving the poles of the bosonic Green function, i.e., $\det\left[\mathcal{D}^{-1}\left(\mathbf{k},\omega_{n}\rightarrow E\left(\mathbf{k}\right)\right)\right]=0$, we obtain the two Bogoliubov spectra,

\begin{widetext}
	\begin{equation}
	E_{\pm}^{2}=\left[\mathcal{A}_{+}\left({\bf k}\right)-\Delta^{2}\right]\pm\sqrt{\mathcal{A}_{-}^{2}\left({\bf k}\right)+\Delta^{2}\Big[\left[C_{1}+D_{1}\text{{\bf k}}^{2}+C_{2}+D_{2}\text{{\bf k}}^{2}\right]^{2}-\left(B_{1{\bf k}}-B_{2{\bf k}}\right)^{2}\Big]},
	\end{equation}
	\end{widetext}
with
$$
\mathcal{A}_{\pm}\left({\bf k}\right)=\frac{\left[B_{1{\bf k}}^{2}-\left(C_{1}+D_{1}\text{{\bf k}}^{2}\right)^{2}\right]\pm\left[B_{2{\bf k}}^{2}-\left(C_{2}+D_{2}\text{{\bf k}}^{2}\right)^{2}\right]}{2}.
$$
Notably, $E_-(\mathbf{k}\to0)=0$, so the lower Bogoliubov branch is gapless, while the upper branch has a gap.

At last, the LHY thermodynamic potential $\Omega_{\text{LHY}}^{\left(0\right)}/\mathcal{V}=-k_{B}T\ln \mathcal{Z}_{2}=\left( k_{B}T/2\mathcal{V}\right)\sum_{\mathbf{k},n}\ln\det\left[-\mathcal{D}^{-1}\left(\mathbf{k},\omega_{n}\right)\right] $, can be obtained from $\mathcal{Z}_{2}=e^{-\frac{1}{\hbar}\int_{0}^{\beta\hbar}\int_{\mathcal{V}}d\mathbf{r}S_{2}}$ within one-loop approximation at zero temperature. After summarizing $\omega_{n}$ in $\sum_{\mathbf{k},n}\ln\det\left[-\mathcal{D}^{-1}\left(\mathbf{k},\omega_{n}\right)\right]$, we explicitly obtain
\begin{equation}
\label{omegaLHY}
\frac{\Omega_{\text{LHY}}^{\left(0\right)}}{\mathcal{V}}
=\frac{1}{2\mathcal{V}}\sum_{\mathbf{k}}\left[E_{+}\left(\mathbf{k}\right)+E_{-}\left(\mathbf{k}\right)\right].
\end{equation}

In the case of equal intraspecies interactions, specifically, $a_{\text{s}_{11}} = a_{\text{s}_{22}} = a_{\text{s}}$ and $r_{\text{eff}}^{\text{s}_{11}} = r_{\text{eff}}^{\text{s}_{22}} = r_{\text{eff}}^{\text{s}}$—we exploit this symmetry to impose uniform condensate parameters: $\mu_1 = \mu_2 = \mu$ and $\phi_{1\text{c}} = \phi_{2\text{c}}$. This consistency condition reduces the coupling constants to $C_1 = C_2 = \mu + \Delta$, $D_1 = D_2 = D$, and $B_{1\mathbf{k}} = B_{2\mathbf{k}} = \varepsilon_{\mathbf{k}} + \mu + 2\Delta + D\mathbf{k}^2$. For the Bogoliubov dispersion relations, this symmetry further simplifies the lower and upper branches to  
$E_{-}\left({\mathbf{k}}\right)=\sqrt{\varepsilon_{\mathbf{k}}\left(\varepsilon_{\mathbf{k}}+2\Delta+2C+2D\text{{\bf k}}^{2}\right)}$ and $E_{+}\left({\mathbf{k}}\right)=\sqrt{\left(\varepsilon_{\mathbf{k}}+2C+2D\text{{\bf k}}^{2}\right)\left(\varepsilon_{\mathbf{k}}+2\Delta\right)}$
respectively.
Thus, the integral in Eq.~(\ref{omegaLHY}) can be separated into two parts $\mathcal{I}_{\pm}$,
	\begin{eqnarray}
		\label{I+-}
\mathcal{I}_{\pm}&=&\frac{1}{2\left(2\pi\right)^{3}}4\pi\int_{0}^{\infty}E_{\pm}k^{2}dk\nonumber\\
&=&\frac{8m^{3/2}}{15\pi^{2}\hbar^{3}}\frac{1}{\left(1+\frac{4mD}{\hbar^{2}}\right)^{2}}C^{5/2}h_{\pm},
	\end{eqnarray}
where $h_{-}=\left(1+\alpha\right)^{5/2}$ and
\begin{equation}
h_{+}=\frac{15}{4}\int_{0}^{\infty}\left(t\right)^{1/2}\sqrt{\left(t+1\right)\left[t+\left(1+\frac{4mD}{\hbar^{2}}\right)\alpha\right]}dt,\nonumber
\end{equation}
with $\alpha=\Delta/C=\Delta/(\mu+\Delta)$ and $D=a_{\text{s}}r_{\text{eff}}^{\text{s}}C/2=a_{\text{s}}r_{\text{eff}}^{\text{s}}\left(\mu+\Delta\right)/2$~\cite{Lorenzi2023} (see
Appendix \ref{A} for a more comprehensive derivation).

Equation~(\ref{I+-}) is one of the key results which embodies the analytical expressions for nonuniversal LHY thermodynamic potential of a weakly interacting bosonic mixtures, with finite-range effects taken into consideration. 
Our results of Eq.~(\ref{I+-}), demonstrate its versatility by successfully recovering previously established results under specific limiting conditions: 
(i) In the absence of the finite-range effects (i.e., $r_{\text{eff}}^{\text{s}}$=0), our results align perfectly with the
Refs.~\cite{Hu2020,Hu2020b,Hu2025}, as verified through the 
parameter $C$, $D$ and $\alpha$. In details, with the condition of $D=a_{\text{s}}r_{\text{eff}}^{\text{s}}C/2=0$, $h_{+}$ recovers to $h_{+}=\left( 15/4\right)\int_{0}^{\infty}\left(t\right)^{1/2}\sqrt{\left(t+1\right)\left(t+\alpha\right)}dt$.
(ii) Then, when the finite-range effects are taken into account (i.e., $r_{\text{eff}}^{\text{s}}\neq0$), our results maintain its consistency with the
Ref.~\cite{Emerson2025} for the reason that $h_{+}\simeq0$ numerically. Notably, $\Omega_{\text{LHY}}^{\left(0\right)}/\mathcal{V}=\left[8m^{3/2}/15\pi^{2}\hbar^{3}\left(1+\frac{4mD}{\hbar^{2}}\right)^{2} \right]C^{5/2}\left(1+\alpha\right)^{5/2}$ in the limit of $\alpha \sim 1$.
Crucially, our approach offers a significant enhancement in self-consistency.

Summing the contributions from the mean-field thermodynamic potential [Eq.~(\ref{omega0})] and the LHY quantum fluctuation correction [Eq.~(\ref{omegaLHY})], we obtain the total thermodynamic potential of the system:  
$\Omega\left(\mu,\Delta\right)/\mathcal{V}=\Omega_0/\mathcal{V} + \Omega_{\text{LHY}}^{(0)}/\mathcal{V}$,
where the explicit form of this total potential is derived in Eq.~(\ref{Omega/V}) in Sec.~\ref{3}. This total potential serves as the key input for computing the nonuniversal EOS of QDs, which is rigorously analyzed in Sec.~\ref{3}.
\section{NONUNIVERSAL EQUATION OF STATE: CHEMICAL POTENTIAL AND ENERGY PER PARTICLE\label{3}}
In the preceding Sec.~\ref{2}, we have delineated the framework
of the bosonic pairing theory. Moving forward, in Sec.~\ref{3},
our objective is to derive the explicit analytical expressions of the self-consistent nonuniversal EOS of the 3D bosonic mixtures, utilizing the thermodynamic potential $\Omega\left(\mu,\Delta\right)/\mathcal{V}$ within one-loop approximation.
The starting point
for this endeavor is to deduce the $\Omega\left(\mu,\Delta\right)/\mathcal{V}=\Omega_0/\mathcal{V} + \Omega_{\text{LHY}}^{(0)}/\mathcal{V}$ combining Eqs.~(\ref{omega0}) and (\ref{omegaLHY}) together as
\begin{widetext}		
\begin{equation}
	\label{Omega/V}
\frac{\Omega}{\mathcal{V}}=-\frac{m}{4\pi\hbar^{2}}\left[\frac{\left(\mu+\Delta\right)^{2}}{a_{\text{s}}}+\frac{\Delta^{2}}{a_{12}}\right]+\frac{8m^{3/2}}{15\pi^{2}\hbar^{3}}\frac{\left(\mu+\Delta\right)^{5/2}}{\left[1+\frac{2m}{\hbar^{2}}a_{\text{s}}r_{\text{eff}}^{\text{s}}\left(\mu+\Delta\right)\right]^{2}}\mathcal{G}\left(\left[1+\frac{2m}{\hbar^{2}}a_{\text{s}}r_{\text{eff}}^{\text{s}}\left(\mu+\Delta\right)\right],\frac{\Delta}{\mu+\Delta}\right).
\end{equation}
\end{widetext}
In what follows, we proceed to obtain the energy per particle of $E/N$ based on Eq.~(\ref{Omega/V}). In more details, we should replace the $\Delta$ and $\mu$ with the particle density of $n$. 

%For a given chemical potential $\mu$, determining the ground-state requires minimizing the thermodynamic potential $\Omega$ in Eq.~(\ref{Omega/V}) to obtain the pairing order parameter $\Delta_0$, followed by calculating the total density via $n = -\partial\Omega/\partial\mu$.  

\textbf{(i) Obtaining the pairing parameter of $\Delta_0\left(\mu \right) $:}

As the first step, we need to obtain the analytical expression of $\Delta$ in Eq.~(\ref{Omega/V}) by $\partial\Omega/\partial \Delta|_{\Delta=\Delta_0}=0 $.   
The mean-field potential $\Omega_0$ in Eq.~(\ref{Omega/V}) contains dominant terms proportional to the interaction parameters $a_{\text{s}}$ and $a_{12}$, which are parametrically larger than the LHY quantum fluctuation correction $\Omega_{\text{LHY}}$, motivated by Refs.~\cite{Hu2020,Hu2020b,Hu2025}. Consequently, when minimizing $\Omega$ with respect to the order parameter $\Delta$, we retain only the $\Omega_0$ contribution, which yields 
\begin{equation}
	\label{mu<<delta0}
	\frac{\Delta_0}{\mu} = -\frac{a_{12}}{{a_{12}}+a_{\text{s}}}.
\end{equation}
From Eq. (\ref{mu<<delta0}), we can immediately notice that the $\Delta_0$ will diverge as $a_{12} \approx -a_{\text{s}}$ and therefore is a non-perturbative result.  This is highly contrast with the mean-field theory in Ref. \cite{Petrov2015}, which can be regarded as the perturbative theory by treating  $\delta a=a_{12}+a_{\text{s}}$ as the small parameter.
Given the characteristic resonant interspecies interaction $a_{12} \approx -a_{\text{s}}$, solving for $\mu$ gives $\mu \ll \Delta_0$,  
	reflecting the hierarchy that the chemical potential $\mu$ is subdominant to the pairing gap in the self-consistent regime.

\textbf{(ii) Obtaining the thermodynamic relation $\mu\left( n\right) $:} 

As the second step, we further obtain the analytical expression of $\mu$ in Eq.~(\ref{Omega/V}) by $n=-\partial\Omega/\partial \mu$.
Given the small magnitude of $|\mu|$ [established in Eq.~(\ref{mu<<delta0})], it is justified to neglect the $\mu$-dependence in the LHY quantum fluctuation correction $\Omega_{\text{LHY}}$ and the $\mu^2$ term in the mean-field potential $\Omega_0$ in Eq.~(\ref{Omega/V}). Under this approximation, the total thermodynamic potential simplifies to $\Omega \approx \Omega_0^{\text{(reduced)}} + \Omega_{\text{LHY}}^{\text{(const)}}$, where $\Omega_0^{\text{(reduced)}}$ retains only $\Delta$-dependent and $\mu^{1}$ terms while $\Omega_{\text{LHY}}^{\text{(const)}}$ is $\mu$-independent. Evaluating $\Omega$ at the saddle point $\Delta = \Delta_0$ (where the pairing order parameter minimizes the potential), we differentiate the simplified expression with respect to $\mu$:  
$n = -\partial \Omega/\partial \mu = m\Delta_0/2\pi\hbar^2 a_{\text{s}} .
$ 
Hence, substituting $\Delta_0$ in Eq.~(\ref{mu<<delta0}) with $\Delta_0\left(n \right) $, we obtain $\mu\left( n\right) $ as
\begin{equation}
	\label{ndel0}
\mu=-\frac{{a_{12}}+a_{\text{s}}}{a_{12}}\frac{2\pi\hbar^2 a_{\text{s}}}{m}n.
\end{equation} 
Equation~(\ref{ndel0}) reveals a divergent chemical potential as \(a_{12} \to 0\) due to its non-perturbative nature, fundamentally differing from one in Ref.~\cite{Petrov2015}. The derivative isolates the dominant mean-field contribution, while quantum fluctuations—encoded in the \(\mu\)-independent $\Omega_{\text{LHY}}^{\text{(const)}}$ term—become negligible in this regime.

\textbf{(iii) Obtaining the per particle energy of $E/N$:} 
 
Finally, following the determination of $\Delta_0(\mu)$ in Eq.~(\ref{mu<<delta0}) and $\mu(n)$ in Eq.~(\ref{ndel0}), we eliminate the chemical potential $\mu$ and pairing order parameter $\Delta_0$ in favor of the total density $n$. Using the self-consistent relation $\mu \ll \Delta_0$ [from Eq.~(\ref{mu<<delta0})] and the expression for $n$ above, we substitute $\Delta_0 = \left( 2\pi\hbar^2 a_{\textsc{s}}/m\right)  n$ into the total energy density $\mathcal{E} = \Omega/\mathcal{V} + \mu n$.  After algebraic simplification, the zero-temperature per-particle energy of the 3D bosonic mixtures is obtained as  
\begin{widetext}
\begin{equation}
	\label{E/N}
\frac{E}{N}=\frac{\Omega}{\mathcal{V}n}+\mu=-\frac{\pi\hbar^{2}}{m}\left(a_{\text{s}}+\frac{a_{\text{s}}^{2}}{a_{12}}\right)n+\frac{32\sqrt{2}\sqrt{\pi}}{15}\frac{\hbar^{2}a_\text{s}^{5/2}}{m}\frac{1}{\left(1+4\pi\frac{r_{\text{eff}}^{\text{s}}}{a_{\text{s}}}na_{\text{s}}^{3}\right)^{2}}\mathcal{G}\left(\frac{r_{\text{eff}}^{\text{s}}}{a_{\text{s}}},na_{\text{s}}^{3}\right)n^{3/2},
\end{equation}
\end{widetext}
with
$
	\mathcal{G}\equiv 4\sqrt{2}+ \frac{15}{4} \int_{0}^{\infty}dt\sqrt{t\left(t+1\right)\left(t+1+4\pi \left( r_{\text{eff}}^{\text{s}}/a_{\text{s}}\right)  na_{\text{s}}^{3}\right)}
$. 

In Equation~(\ref{E/N}), the first two terms constitute the mean-field component, corresponding to the condensate atoms, which is not affected by the finite-range interaction, while the
subsequent terms represent the excitation part stemming from the excited atoms. 
Note that when the system approaches the collapse phase ($\delta a=a_{12}+a_{\text{s}}<0$, $a_{12}<0$), the LHY term in Eq.~(\ref{E/N}) becomes mathematically well-defined.
We remark that in the absence of the finite-range interaction (i.e., $r_{\text{eff}}^{\text{s}}=0$), $\mathcal{G}=4\sqrt{2}+\left( 15/4\right) \int_{0}^{\infty}dt\sqrt{t}\left(t+1\right)=4\sqrt{2}$ [see Eq.~(\ref{integralG}) in Appendix~\ref{A} for more calculation details], resulting in  energy per particle in Eq.~(\ref{E/N})
\begin{equation}
\frac{E}{N}=-\frac{\pi\hbar^{2}}{m}\left(a_{\text{s}}+\frac{a_{\text{s}}^{2}}{a_{12}}\right)n+\frac{256\sqrt{\pi}}{15}\frac{\hbar^{2}a_\text{s}^{5/2}}{m}n^{\frac{3}{2}}.\label{E/N2}
 \end{equation}
 We demonstrate that Eq.~(\ref{E/N2}) reproduces Eq.~(11) of Ref.~\cite{Hu2020} exactly. By comparing with Petrov's original result, we find that the approximate LHY correction adopted in Ref.~\cite{Petrov2015} is naturally recovered by our pairing theory and becomes exact in the special case of $\left|a_{\text{s}}\right|= \left|a_{12}\right|$ $(\delta a\simeq0)$. Notably, the mean-field energy term (first term in Eqs.~(\ref{E/N}) and (\ref{E/N2})) is modified by a factor of $-a_{\text{s}}/a_{12}$ ($a_{12} < 0$), which quantitatively explains the discrepancy between theory and experiment observed in quantum droplet equilibrium densities  and critical particle numbers firstly noted by Ref.~\cite{Hu2020}. This result highlights the essential role of pairing correlations in stabilizing self-bound quantum droplets beyond mean-field predictions.  When we take finite-range interaction into account (i.e., $r_{\text{eff}}^{\text{s}}\neq0$), the LHY correction becomes interesting for a divergency in Eq.~(\ref{E/N}) when $1+4\pi\left(r_{\text{eff}}^{\text{s}}/a_{\text{s}} \right)na_{\text{s}}^{3}=0$, indicating a singular phenomenon in nonuniversal EOS~\cite{Emerson2025,Ye2026}.

%We remark that Eq. (\ref{E/N2}) can exactly recover Eq. (11) in Ref.~\cite{Hu2020}. By Eq. (\ref{E/N2}) with the corresponding Petrov's result in Ref. \cite{Petrov2015}, we note  that the
%approximate LHY term adopted by Petrov is reproduced by our pairing theory and is actually exact at the special case of $a=a_{12}$. In contrast, the mean-field energy, the first term in
%Eqs. (\ref{E/N}) and (\ref{E/N}), is now changed by a factor of $-a/a_{12}<1$.

Another crucial EOS for bosonic mixtures is characterized by the equilibrium density $ n_{\text{eq}}$ where the energy per particle $E/N$ attains its minimum value. To determine this critical density, we employ the variational principle by taking the functional derivative of the energy density expression in Eq.~(\ref{E/N}) with respect to particle density $ n $:
\begin{equation}
	\left. \frac{\partial}{\partial n} \left( \frac{E}{N} \right) \right|_{n = n_{\text{eq}}} = 0.\nonumber
\end{equation}
The analytical derivation of $n_{\text{eq}}$ encounters a fundamental challenge due to the non-integrable singularity in the interaction term of Eq.~(\ref{E/N}), which prevents explicit solution for $n_{\text{eq}}$. However, we develop an alternative analytical framework by considering the dimensionless gas parameter $\bar{n} = n a_{\text{s}}^3$. This approach yields an implicit equation for the equilibrium condition:
\begin{widetext}
	\begin{eqnarray}
		\label{derivative E/N}
		\frac{\partial E/N}{\partial\left(na_{\text{s}}^{3}\right)}=&-&2\pi\left(1+\frac{a_{\text{s}}}{a_{12}}\right)\frac{\hbar^{2}}{2ma_{\text{s}}^{2}}+\frac{32\sqrt{2}\sqrt{\pi}}{5}\left(na_{\text{s}}^{3}\right)^{1/2}\frac{1}{\left(1+4\pi\frac{r_{\text{eff}}^{\text{s}}}{a_{\text{s}}}na_{\text{s}}^{3}\right)^{2}}\mathcal{\mathcal{G}}\left(\frac{r_{\text{eff}}^{\text{s}}}{a_{\text{s}}},na_{\text{s}}^{3}\right)\frac{\hbar^{2}}{2ma_{\text{s}}^{2}}\nonumber\\
		&-&\frac{512\sqrt{2}\pi^{3/2}}{15}\frac{r_{\text{eff}}^{\text{s}}}{a_{\text{s}}}\left(na_{\text{s}}^{3}\right)^{3/2}\frac{1}{\left(1+4\pi\frac{r_{\text{eff}}^{\text{s}}}{a_{\text{s}}}na_{\text{s}}^{3}\right)^{3}}\mathcal{\mathcal{G}}\left(\frac{r_{\text{eff}}^{\text{s}}}{a_{\text{s}}},na_{\text{s}}^{3}\right)\frac{\hbar^{2}}{2ma_{\text{s}}^{2}}\nonumber\\
		&+&32\sqrt{2}\pi^{3/2}\frac{r_{\text{eff}}^{\text{s}}}{a_{\text{s}}}\left(na_\text{s}^{3}\right)^{3/2}\frac{1}{\left(1+4\pi\frac{r_{\text{eff}}^{\text{s}}}{a_{\text{s}}}na_{\text{s}}^{3}\right)^{2}}\mathcal{F}\left(\frac{r_{\text{eff}}^{\text{s}}}{a_{\text{s}}},na_{\text{s}}^{3}\right)\frac{\hbar^{2}}{2ma_{\text{s}}^{2}},
	\end{eqnarray}
\end{widetext}
with
$\mathcal{F}=\int_{0}^{\infty}dt\sqrt{t\left(t+1\right)}/\sqrt{t+1+4\pi \left( r_{\text{eff}}^{\text{s}}/a_{\text{s}}\right) na_{\text{s}}^{3}}$.
We systematically investigate the energy per particle $E/N$ defined in Eq.~(\ref{E/N}) and its density derivative $\partial\left(E/N \right)/\partial (na_{\text{s}}^{3})$ from Eq.~(\ref{derivative E/N}) as functions of particle density $ n $. This comparative analysis reveals the critical role of finite-range interactions in shaping the EOS for QDs. By mapping the density dependence of these quantities, we demonstrate how the balance between mean-field attraction and quantum fluctuation-induced repulsion determines the stability window of self-bound droplets. 

Before proceeding to the detailed analysis of the density dependence of the energy per particle $E/N$ and its derivative as described in Eqs.~(\ref{E/N}) and (\ref{derivative E/N}), 
it is instructive to examine the rationality of the dimensionless parameters central to our model: the finite-range coupling constant $r^{\text{s}}_{\text{eff}}/a_{\text{s}}$
and the gas parameter $na_{\text{s}}^{3}$. This preliminary assessment is essential for establishing the experimental viability of our approach. As a concrete example, we consider a quasi-BEC of $^{39}\text{K}$. For the scattering length $a_{\text{s}}$, we adopt the typical value $a_{\text{s}}\approx34.6 a_{0}$
from the Feshbach resonance data compiled in Ref.~\cite{Chin2010} with $a_{0}$ being Bohr radius. For the effective range $r^{\text{s}}_{\text{eff}}$, we utilize the experimentally constrained values (e.g., $r^{\text{s}}_{\text{eff}}\approx -1158.9 a_{0}$
or $1021.2a_{0}$) reported for the same system $^{39}\text{K}$ in Ref.~\cite{Cikojevi2020,Cikojevi2020b}. Crucially, according to the dual-laser ``molecular dark-state'' scheme proposed in Ref.~\cite{Wu2012,Haibin2012}, we can independently tune $a_{\text{s}}$ and $r^{\text{s}}_{\text{eff}}$
without being constrained by the intrinsic properties of a single interaction potential. Adhering to the Wigner causality bound ($|r^{\text{s}}_{\text{eff}}|\leq 2r_{0}$, where $r_{0}$ is the van der Waals radius), we can adjust $a_{\text{s}}$ via magnetic Feshbach resonance to approach zero while keeping $r^{\text{s}}_{\text{eff}}$ fixed at its physical value. This allows us to reach the extreme regimes of $r^{\text{s}}_{\text{eff}}/a_{\text{s}}=\pm 200$. For a typical density of $n\approx4\times10^{13}\text{cm}^{-3}$~\cite{Ferioli2020}, the resulting gas parameter is $n a_{\text{s}}^{3}\sim 10^{-5}$. These parameter ranges affirm the physical relevance and experimental accessibility of our model.

%Before proceeding to the detailed analysis of the density dependence of the energy per particle of $E/N$ and its derivative as described in Eqs.~(\ref{E/N}) and (\ref{derivative E/N}), it is instructive to examine the rationality of the dimensionless parameters central to our model: the finite-range coupling constant $r_{\text{eff}}^{\text{s}}/a_{\text{s}}$ and the gas parameter $na_{\text{s}}^{3}$. This preliminary assessment is essential for establishing the experimental viability of our approach. 
%As a concrete example, we consider $^{6}$Li, following Ref.~\cite{Bartenstein2005}. In this context, the typical density is $n \approx 4 \times 10^{13}~\text{cm}^{-3}$, and the scattering length is $a_{\text{s}} \approx 1.13 \times 10^{-8}~\text{m}$. This yields the magnitude of gas parameter $na_{\text{s}}^{3} \sim 10^{-5}$. Furthermore, according to Ref.~\cite{Wu2012}, the effective range $r_{\text{eff}}^{\text{s}}$ is estimated to lie within $0$ to $\pm 3.71 \times 10^{-6}~\text{m}$. 
%Substituting these values, we find $r_{\text{eff}}^{\text{s}}/a_{\text{s}} \in [0, \pm 200]$ and $na_{\text{s}}^{3} \sim 10^{-5}$. These parameter ranges affirm the physical relevance and experimental accessibility of our model.

\begin{figure}[hbtp] 
	 \begin{centering}
		\includegraphics[scale=0.75]{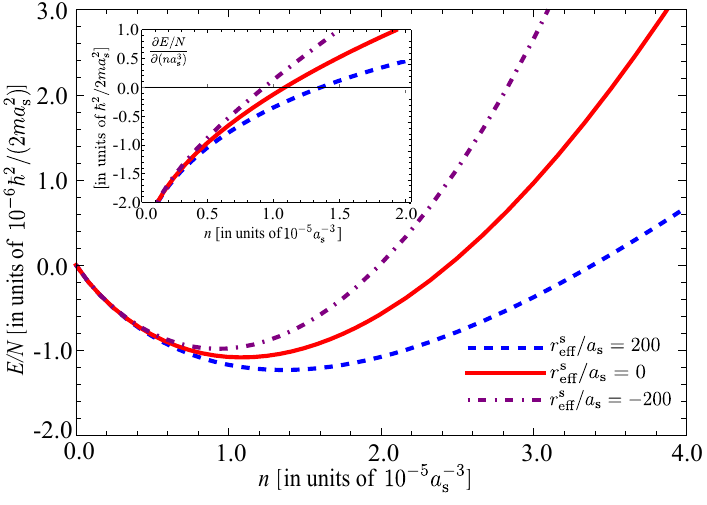} 
		\par\end{centering} 
	\caption{
		Energy per particle, $E/N$, as a function of particle density $n$ for an interspecies scattering length $a_{12} = -1.05a_{\text{s}}$ and equal intraspecies lengths $a_{\text{s}_{11}} = a_{\text{s}_{22}} \equiv a_{\text{s}}$. The purple dot-dashed and blue dashed curves correspond to finite-range interactions with $r_{\text{eff}}^{\text{s}}/a_{\text{s}} = -200$ and $200$, respectively, while the red solid curve represents the zero-range universal prediction by Hui Hu \cite{Hu2020,Hu2020b,Hu2025} ($r_{\text{eff}}^{\text{s}}/a_{\text{s}} = 0$). The inset shows the derivative $\partial(E/N)/\partial na_{\text{s}}^3$, emphasizing the shift in equilibrium density $n_{\text{eq}}$ due to finite-range effects. 
		\label{fig1}
	}
\end{figure}

We now present a comprehensive analysis of finite-range interaction effects on the energy per particle $ E/N $ and equilibrium density $ n_{\text{eq}} $ in bosonic mixtures. As illustrated in Figure~\ref{fig1}, these effects manifest through the density dependence of $ E/N $ and its derivative, derived from Eqs.~(\ref{E/N}) and (\ref{derivative E/N}), under interspecies interaction $ a_{12} = -1.05a_{\text{s}} $ with equal intraspecies interactions $ a_{\text{s}_{11}} = a_{\text{s}_{22}} \equiv a_{\text{s}} $. The calculations incorporate finite-range corrections parameterized by $ r_{\text{eff}}^{\text{s}}/a_{\text{s}} $.
Our results of Eqs.~(\ref{E/N}) and (\ref{derivative E/N}), demonstrate their versatility by successfully recovering previously established results under specific limiting conditions: 
(i) In the absence of the finite-range effects (i.e., $r_{\text{eff}}^{\text{s}}$=0), we reproduce the zero-range result of Refs.~\cite{Hu2020,Hu2020b,Hu2025}, as verified through the the red solid curve.
(ii) Meanwhile, the purple dot-dashed and blue dashed curves in Figure~\ref{fig1} demonstrate how the attractive ($r_{\text{eff}}^{\text{s}}/a_{\text{s}}=-200$) and
repulsive ($r_{\text{eff}}^{\text{s}}/a_{\text{s}}=200$) finite-range interaction can affect the $E/N$ and $n_{\text{eq}}$. 

As shown in Figure~\ref{fig1}, the finite-range interaction has relative small effects on energy per particle of
$E/N$ compared with the case of vanishing the finite-range
interaction when the gas parameter of $na_{\text{s}}^{3}$ is small. 
In contrast, with the increase of the $na_{\text{s}}^{3}$, the effect of finite-range
interaction on energy per particle of
$E/N$ and equilibrium density of $n_{\text{eq}}$ becomes significant.
For example, at the typical experimental parameter onset of $na_{\text{s}}^{3}\sim 3\times 10^{-5}$~\cite{Chin2010,Ferioli2020}, the derivation of the energy per particle of
$E/N$ in case of $r_{\text{eff}}^{\text{s}}/a_{\text{s}}=200$ from the case of $r_{\text{eff}}^{\text{s}}/a_{\text{s}}=-200$ is
calculated as more than $3\times10^{-6}$ in units of $\hbar^{2}/2m a_{\text{s}}^{2}$, while the dimensionless equilibrium density of $n_{\text{eq}}a_{\text{s}}^{3}$ in inset in Figure~\ref{fig1} in case of $r_{\text{eff}}^{\text{s}}/a_{\text{s}}=200$ from the case of $r_{\text{eff}}^{\text{s}}/a_{\text{s}}=-200$ is about $10^{-5}$, showing that the finite-range effect is well in reach in recent experiments through magnetic and optical Feshbach resonances~\cite{Chin2010,Haibin2012,Wu2012,Inouye1998,Tanzi2018,Knoop2011}.

The chemical potential $\mu$, a key thermodynamic quantity for the bosonic mixtures, is defined as $\mu = \partial \mathcal{E} / \partial n$, where $\mathcal{E} = E / \mathcal{V}$ denotes the energy density. From $\mu$, the inverse compressibility $\kappa^{-1}=n\frac{\partial \mu}{\partial n}$ follows directly. This allows us to examine the speed of sound, $c_{\text{s}} = \sqrt{1 / (\kappa m)}$, which encodes nonuniversal quantum effects. Thus, based on the Eq.~(\ref{E/N}), we can analytically obtained the chemical potential $\mu$ as
\begin{eqnarray}
	\label{mu}
\!\!\!\!\!\!\!\!\!\!\!\mu&=&g_{0}n\Big\{-\frac{1}{2}\left(1+\frac{a_{\text{s}}}{a_{12}}\right)\nonumber\\
\!\!\!\!\!\!\!\!&+&\frac{4\sqrt{2}}{3\sqrt{\pi}}\frac{1}{\left(1+4\pi\frac{r_{\text{eff}}^{\text{s}}}{a_{\text{s}}}na_{\text{s}}^{3}\right)^{2}}\mathcal{G}\left(\frac{r_{\text{eff}}^{\text{s}}}{a_{\text{s}}},na_{\text{s}}^{3}\right)\left(na_{\text{s}}^{3}\right)^{1/2}\nonumber\\
\!\!\!\!\!\!\!\!&-&\frac{64\sqrt{2\pi}}{15}\frac{r_{\text{eff}}^{\text{s}}}{a_{\text{s}}}\frac{1}{\left(1+4\pi\frac{r_{\text{eff}}^{\text{s}}}{a_{\text{s}}}na_{\text{s}}^{3}\right)^{3}}\mathcal{G}\left(\frac{r_{\text{eff}}^{\text{s}}}{a_{\text{s}}},na_{\text{s}}^{3}\right)\left(na_{\text{s}}^{3}\right)^{3/2}\nonumber\\
\!\!\!\!\!\!\!\!&+&4\sqrt{2\pi}\frac{r_{\text{eff}}^{\text{s}}}{a_{\text{s}}}\frac{1}{\left(1+4\pi\frac{r_{e}}{a}na^{3}_{\text{s}}\right)^{2}}\mathcal{F}\left(\frac{r_{\text{eff}}^{\text{s}}}{a_{\text{s}}},na_{\text{s}}^{3}\right)\left(na_{\text{s}}^{3}\right)^{3/2}\Big\}.
\end{eqnarray}

In Equation~(\ref{mu}), $g_0$ denotes the intraspecies zero-range coupling constant, while $r_{\text{eff}}^{\text{s}}/a_{\text{s}}$ and $n a_{\text{s}}^3$ represent the dimensionless finite-range coupling constant and the gas parameter. The integrals $\mathcal{G}(r_{\text{eff}}^{\text{s}}/a_{\text{s}}, n a_{\text{s}}^3)$ and $\mathcal{F}(r_{\text{eff}}^{\text{s}}/a_{\text{s}}, n a_{\text{s}}^3)$ have been defined in Eqs.~(\ref{E/N}) and (\ref{derivative E/N}). We analyze the effect of the nonuniversal LHY correction on the chemical potential $\mu$, as obtained from the bosonic pairing theory in Sec.~\ref{4}.

We remark that, utilizing the consistent field theory within the framework of HS transformation, we have derived key analytical expressions for the nonuniversal EOS for a 3D bosonic mixtures at zero temperature. Specifically, our results yield explicit density-functional relationships for the energy per particle of $E/N$ and equilibrium density of $n_{\text{eq}}$, incorporating finite-range interaction effects through a renormalized pseudopotential approach.  
This development sheds new light on modeling QDs, where precise characterization of short-range interatomic correlations remains crucial for understanding superfluid behavior and metastable states observed in recent helium nanodroplet experiments. 
\section{FREQUENCY SHIFTS OF THE BOSONIC MIXTURE\label{4}}

In Sec.~\ref{3}, we employ consistent field theory within the HS transformation framework to rigorously derive the nonuniversal LHY correction to the EOS~[see Eqs. (\ref{E/N}), (\ref{derivative E/N}) and (\ref{mu})] for a bosonic mixtures with equal intraspecies interactions $a_{\text{s}_{11}} = a_{\text{s}_{22}} \equiv a_{\text{s}}$. The purpose of Sec.~\ref{4} is to propose an experimental protocol to observe these nonuniversal LHY corrections to the chemical potential $\mu$ via frequency shifts in the breathing modes~\cite{Hu2011,Liang2010,Yin2020}.

In a two-component quantum gas, there are two excitation channels: the total density channel, associated with $n(\mathbf{r},t)=n_1(\mathbf{r},t)+n_2(\mathbf{r},t)$, and the spin-density channel, linked to $S_z(\mathbf{r},t)=n_1(\mathbf{r},t)-n_2(\mathbf{r},t)$. Previous work~\cite{Wu2018} on spinor Bose gases has showed that spin-density fluctuations (the two-component internal degrees of freedom) decouple from density fluctuations in determining excitation frequencies. Thus, the two-component attractive case studied here is equivalent to a single-component description with proper renormalization of the scattering length~\cite{Science2018}. Following the standard methodology outlined in Refs.~\cite{Stringari1996, Pitaevskii1998,Hu2020c,Zhang2023,Mukherjee2025,Tylutki2020,S.V.2025}, we formulate the hydrodynamic equation~\cite{Dalfovo1999,Hu2014} as follows:
\begin{equation}
	\label{hydrodynamic equation}
	m\frac{\partial^{2}\delta n\left(\mathbf{r},t \right) }{\partial t^{2}}-\nabla\cdot\left[n\left(\mathbf{r} \right) \nabla\left(\frac{\partial \mu_{l}}{\partial n}\delta n\left(\mathbf{r},t \right) \right)  \right]=0, 
\end{equation}
where $n\left(\mathbf{r}\right) =n_{1}\left(\mathbf{r}\right)+n_{2}\left(\mathbf{r}\right)$, $\delta n=\delta  n_{1}+\delta n_{2}$.

As the first step, using Eq.~(\ref{mu}), for the bosonic mixtures with interspecies interaction $a_{\text{s}}/a_{12} = -1/1.05$ and equal intraspecies interactions $a_{\text{s}_{11}}=a_{\text{s}_{22}}\equiv a_{\text{s}}$, we derive the ground-state density expansion as
\begin{widetext}
\begin{equation}
	\label{n}
n\left(\mathbf{r}\right)=n_{\text{TF}}\left[1-\frac{8\sqrt{2}}{3\sqrt{\pi}\left(1+\frac{a_{\text{s}}}{a_{12}}\right)}\frac{1}{\left(1+4\pi\frac{r_{\text{eff}}^{\text{s}}}{a_{\text{s}}}n_{\text{TF}}a_{\text{s}}^{3}\right)^{2}}\mathcal{G}\left(\frac{r_{\text{eff}}^{\text{s}}}{a_{\text{s}}},n_{\text{TF}}a_{\text{s}}^{3}\right)\left(n_{\text{TF}}a_{\text{s}}^{3}\right)^{1/2}\right].
\end{equation}
\end{widetext}
In Equation.~(\ref{n}), $n_{\text{TF}}=\left( g_{0}n-V_{\text{ext}}\right) /g_{0}$~\cite{Stringari1996,Pitaevskii1998}, is the so-called Thomas-Fermi (TF) result for the ground-state density, while the dimensonless integral $\mathcal{G}\left(r_{\text{eff}}^{\text{s}}/a_{\text{s}},n_{\text{TF}}a_{\text{s}}^{3}\right)$ has been definend in Sec.~\ref{3} by substituting $n$ with $n_{\text{TF}}$. Notably, the second term represents the nonuniversal LHY effects induced by finite-range interactions.

Subsequently, we derive the expansion of $\left[n\partial \mu_{l}/\partial n \right]$ as a
series in terms of $n_{\text{TF}}$ by substituting Eq.~(\ref{n}) into Eq.~(\ref{mu}).
Following this substitution, we insert both Eqs. (\ref{mu}) and (\ref{n})
into the Eq.~(\ref{hydrodynamic equation}). Through meticulous algebraic manipulations, we ultimately arrive at
\begin{widetext}
\begin{equation}
	\label{oequation}
	m\omega^{2}\delta n+\nabla\cdot\left[g_{0}n_{\text{TF}}\nabla\delta n\right]=-\frac{1}{2}\nabla^{2}\left[\frac{4\sqrt{2}}{3\sqrt{\pi}}\left(a_{\text{s}}^{3}\right)^{1/2}\frac{1}{\left(1+4\pi\frac{r_{\text{eff}}^{\text{s}}}{a_{\text{s}}}n_{\text{TF}}a_{\text{s}}^{3}\right)^{2}}\mathcal{G}\left(\frac{r_{\text{eff}}^{\text{s}}}{a_{\text{s}}},n_{\text{TF}}a_{\text{s}}^{3}\right)g_{0}n_{\text{TF}}^{3/2}\delta n\right].
\end{equation}
\end{widetext}
In the limit $n_{\text{TF}}a_{\text{s}}^{3}=0$, Eq.~(\ref{oequation}) reduces to the basic hydrodynamic equation
$m\omega^{2}\delta n + \nabla \cdot [g_{0} n_{\text{TF}} \nabla \delta n] = 0$. 
The frequency $\omega$ then admits the analytical form 
$\omega(n_{r},l) = \omega_{0} \sqrt{2n_{r}^{2} + 2n_{r}l + 3n_{r} + l}$, 
where $n_{r}$ and $l$ denote the radial node number and the angular momentum quantum number associated with the excitation. 
This analytical expression provides a direct link between the excitation frequency and its quantum numbers.

Finally, Eq.~(\ref{oequation}) can be solved perturbativly by treating its right-hand side as a small perturbation, from which the analytical expression for the frequency shifts follows:
\begin{widetext}
	\begin{equation}
		\label{doo1}
	\frac{\delta\omega}{\omega}=-\frac{4\sqrt{2}}{3\sqrt{\pi}}\frac{g_{0}}{4m\omega^{2}}\left(a_{\text{s}}^{3}\right)^{1/2}\frac{\int d^{3}{\bf r}\left(\nabla^{2}\delta n^{*}\right)\left[n_{\text{TF}}^{3/2}\delta n\frac{1}{\left(1+4\pi\frac{r_{\text{eff}}^{\text{s}}}{a_{\text{s}}}n_{\text{TF}}a_{\text{s}}^{3}\right)^{2}}\mathcal{G}\left(\frac{r_{\text{eff}}^{\text{s}}}{a_{\text{s}}},n_{\text{TF}}a_{\text{s}}^{3}\right)\right]}{\int d^{3}{\bf r}\delta n^{*}\delta n}.
	\end{equation}
\end{widetext}

The integrals in Eq.~(\ref{doo1}) are evaluated over the region where the TF density is positive. To probe the nonuniversal LHY corrections in QDs, we focus on compressional modes, which are particularly sensitive to changes in the EOS. Among these, the breathing mode in a spherical trap—corresponding to the quantum numbers $(n_{r} = 1,\ l = 0)$—is the fundamental excitation. While, its frequency is given by $\omega=\sqrt{5}\,\omega_{0}$ to zeroth order, and its density oscillation follows $\delta n \sim\left[ r^{2}-\left( 3/5\right) R^{2}\right] $. In this case, Eq.~(\ref{doo1}) yields
\begin{widetext}
	\begin{equation}
		\label{deltaOmega/Omega2}
		\frac{\delta\omega}{\omega}=\frac{2\sqrt{2}}{5\sqrt{\pi}}\frac{\left[n\left(0\right)a_{\text{s}}^{3}\right]^{1/2}}{\left(1+\frac{a_{\text{s}}}{a_{12}}\right)}\frac{\int_{0}^{1}x^{2}dx\left(1-x^{2}\right)^{3/2}\left(x^{2}-\frac{3}{5}\right)\frac{1}{\left(1+4\pi\frac{r_{\text{eff}}^{\text{s}}}{a_{\text{s}}}\left(1-x^{2}\right)n\left(0\right)a_{\text{s}}^{3}\right)^{2}}\left[2^{5/2}+\mathcal{G}\left[\frac{r_{\text{eff}}^{\text{s}}}{a_{\text{s}}},\left(1-x^{2}\right)n\left(0\right)a_{\text{s}}^{3}\right]\right]}{\int_{0}^{1}x^{2}dx\left(x^{2}-\frac{3}{5}\right)^{2}},
	\end{equation}
\end{widetext}
revealing the fractional shift in the breathing-mode frequency,
where $n\left( 0\right) $ represents the density evaluated at the center of the trap.
The dimensionless parameters $r_{\text{eff}}^{\text{s}}/a_{\text{s}}$ and $n(0)a_{\text{s}}^{3}$ have been rationalized in Sec.~\ref{3}.
We are now poised to analyze the impact of the nonuniversal EOS in Eq.~(\ref{mu}), arising from
the finite-range interaction parameter $r_{\text{eff}}^{\text{s}}$, on this fractional frequency shift of the breathing-mode given by Eq.~(\ref{deltaOmega/Omega2}). 
\begin{figure}[htbp] 
	\begin{centering}
		\includegraphics[scale=0.53]{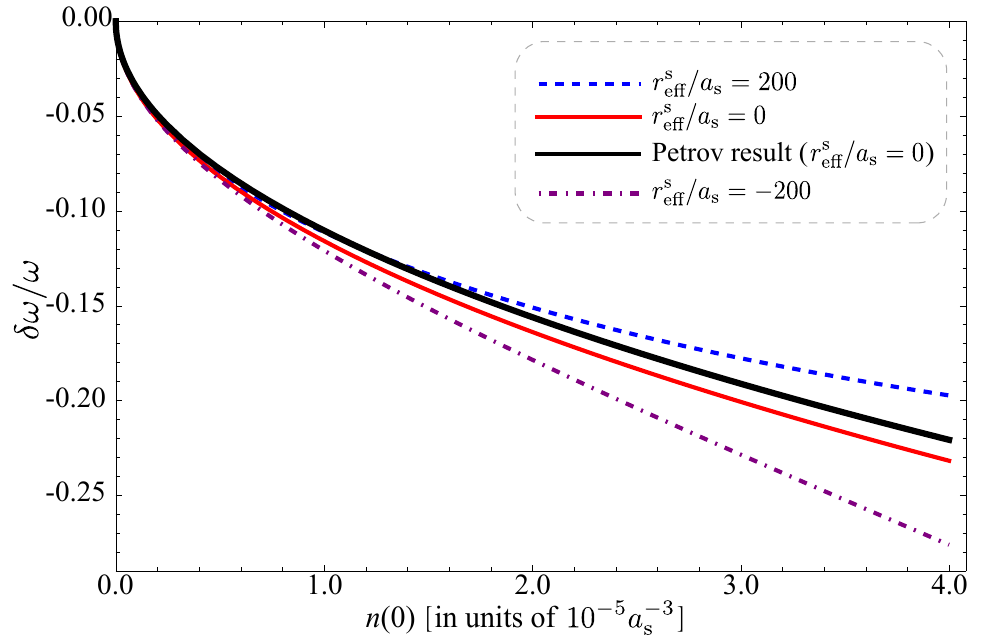} 
		\par\end{centering} 
	\caption{
		Frequency shifts $\delta\omega/\omega$ as a function of the gas parameter for different values of the finite-range coupling $r_{\text{eff}}^{\text{s}}/a_{\text{s}}$, computed from Eq.~(\ref{deltaOmega/Omega2}) for the breathing mode of a 3D bosonic mixtures. The system has interspecies interaction $a_{\text{s}}/a_{12} = -1/1.05$ and equal intraspecies scattering lengths $a_{\text{s}_{11}} = a_{\text{s}_{22}} \equiv a_{\text{s}}$. The black bold-solid curve corresponds to Petrov's universal EOS, while the red solid curve is obtained from the self-consistent bosonic pairing theory. Our nonuniversal results—blue dashed and purple dot-dashed curves—incorporate finite-range interactions, with $r_{\text{eff}}^{\text{s}}/a_{\text{s}} = -200$ (attractive) and $r_{\text{eff}}^{\text{s}}/a_{\text{s}} = 200$ (repulsive), respectively.
		\label{fig2}
	}
\end{figure}

To visualize this effect, we have plotted frequency shifts of the breathing mode $\delta \omega /\omega$ in Eq.~(\ref{deltaOmega/Omega2}) into Figure~\ref{fig2}, showing how the dimensionless finite-range interaction parameter of $r_{\text{eff}}^{\text{s}}/a_{\text{s}}$ can affect the frequency shifts of $\delta \omega /\omega$. The red solid and black bold-solid curves correspond to the zero-range case ($r_{\mathrm{s}}/a_{\mathrm{s}}=0$), obtained respectively from the present self-consistent field theory~\cite{Hu2020,Hu2020b,Hu2025} (HS transformation) and from Petrov's universal EOS, $E/N = \pi\hbar^{2}(a_{\mathrm{s}}+a_{12})n + (256\sqrt{\pi}/15)(\hbar^{2}a_{\mathrm{s}}^{5/2}/m)n^{3/2}$. The blue dashed and purple dot-dashed curves show the effect of a repulsive ($r_{\mathrm{s}}/a_{\mathrm{s}}=200$) and an attractive ($r_{\mathrm{s}}/a_{\mathrm{s}}=-200$) finite-range interaction, calculated within the self-consistent field theory.
The finite-range correction to $\delta\omega/\omega$ is relatively small at low gas parameter $n\left( 0\right) a_{\mathrm{s}}^{3}$ but grows noticeably as $n\left(0 \right) a_{\mathrm{s}}^{3}$ increases. At a typical experimental value $n\left( 0\right) a_{\mathrm{s}}^{3} \sim 4\times10^{-5}$, the shift between the $r_{\mathrm{s}}/a_{\mathrm{s}} = 200$ and $-200$ cases exceeds $5\%$. Given that collective frequencies can already be measured with a precision better than $0.3\%$, this finite-range effect is well within experimental reach, offering a route to probe the nonuniversal LHY corrections in QDs.

\section{CONCLUSION AND OUTLOOK \label{5}}

In summary, we present a unified theoretical framework combining self-consistent field theory with HS transformation to investigate nonuniversal EOS for 3D bosonic mixtures under finite-range interactions. Our analytical solutions for energy per particle and equilibrium density explicitly incorporate self-consistent LHY corrections, bridging quantum fluctuation effects with short-range interatomic potential details. This extends previous studies \cite{Hu2020,Emerson2025,Ye2026} by establishing systematic corrections to nonuniversal LHY EOS regimes, revealing how finite-range interactions modify QDs stabilization through balanced interspecies attraction and 3D quantum pressure.

We further demonstrate that these self-consistent LHY effects produce characteristic frequency shifts in breathing-mode oscillations, with analytical predictions matching experimental observables in current cold-atom setups \cite{Science2018}. Our results establish a paradigm for exploring nonuniversal quantum hydrodynamics beyond mean-field theory, where finite-range interaction details directly manifest in collective excitations and equilibrium properties. Experimental verification of these frequency modulations would constitute a definitive test of quantum fluctuation-dominated stabilization mechanisms in multi-component Bose systems.

\section{ACKNOWLEDGMENTS}

We thank Tao Yu, Ying Hu, and Biao Wu for stimulating discussions and useful help. This work was supported by the
National Natural Science Foundation of China under Grant No. 12574301 and the Zhejiang Provincial Natural Science Foundation of China under Grant No. LZ25A040004.
\appendix
\begin{widetext}
	\section{Detailed derivation of Eq.~(\ref{Omega/V})\label{A}}
For the purpose of maintaining self-consistency within this
work, we provide a concise overview of the key steps of
the derivation of Eq.~(\ref{Omega/V}). 
By substituting the expression of $E_{-}\left(\mathbf{k} \right) $ with $E_{-}\left({\mathbf{k}}\right)=\sqrt{\varepsilon_{\mathbf{k}}\left(\varepsilon_{\mathbf{k}}+2\Delta+2C+2D\text{{\bf k}}^{2}\right)}$ into $\mathcal{I_{-}}$ and complete the integral $\mathcal{I_{-}}$ by Gamma function to avoid ultraviolet divergence
	\begin{eqnarray}
		\label{I-}
		\mathcal{I}_{-}&=&\frac{1}{2\left(2\pi\right)^{3}}4\pi\int_{0}^{\infty}E_{-}k^{2}dk\nonumber\\
		&=&\frac{1}{2\pi ^2}\int_{0}^{\infty}k^{2}\sqrt{\frac{\hbar^{2}k^{2}}{2m}\left(\frac{\hbar^{2}k^{2}}{2m}+2\Delta+2C+2Dk^{2}\right)}dk\nonumber\\
		&=&\frac{1}{8\pi^{2}}\left(\frac{2m}{\hbar^{2}}\right)^{3/2}\left(\frac{m^{*}}{m}\right)^{2}\int_{0}^{\infty}\left(\frac{\hbar^{2}k^{2}}{2m^{*}}\right)^{1/2}\sqrt{\frac{\hbar^{2}k^{2}}{2m^{*}}\left(\frac{\hbar^{2}k^{2}}{2m^{*}}+2\Delta+2C\right)}d\left(\frac{\hbar^{2}k^{2}}{2m^{*}}\right)\nonumber\\
		&=&\frac{1}{8\pi^{2}}\left(\frac{2m}{\hbar^{2}}\right)^{3/2}\left(\frac{m^{*}}{m}\right)^{2}\left[2\left(\Delta+C\right)\right]^{5/2}\int_{0}^{\infty}t\left(1+t\right)^{1/2}dt\nonumber\\
		&=&\frac{1}{8\pi^{2}}\left(\frac{2m}{\hbar^{2}}\right)^{3/2}\left(\frac{m^{*}}{m}\right)^{2}\left[2\left(\Delta+C\right)\right]^{5/2}\frac{\Gamma\left(2\right)\Gamma\left(-\frac{5}{2}\right)}{\Gamma\left(-\frac{1}{2}\right)}\nonumber\\
		&=&\frac{8m^{3/2}}{15\pi^{2}\hbar^{3}}\left(\frac{m^{*}}{m}\right)^{2}C^{5/2}\left(1+\frac{\Delta}{C}\right)^{5/2}\nonumber\\
		&=&\frac{8m^{3/2}}{15\pi^{2}\hbar^{3}}\frac{1}{\left(1+\frac{4mD}{\hbar^{2}}\right)^{2}}C^{5/2}\left(1+\alpha\right)^{5/2},
	\end{eqnarray}
with $\alpha=\Delta/C$.

Meanwhile, by substituting the expression of $E_{+}\left(\mathbf{k} \right)$ with $E_{+}\left({\mathbf{k}}\right)=\sqrt{\left(\varepsilon_{\mathbf{k}}+2C+2D\text{{\bf k}}^{2}\right)\left(\varepsilon_{\mathbf{k}}+2\Delta\right)}$ into $\mathcal{I_{+}}$
	\begin{eqnarray}
		\label{I+}
		\mathcal{I}_{+}&=&\frac{1}{2\left(2\pi\right)^{3}}4\pi\int_{0}^{\infty}E_{+}k^{2}dk\nonumber\\
		&=&\frac{1}{2\pi ^2}\int_{0}^{\infty}k^{2}\sqrt{\left(\frac{\hbar^{2}k^{2}}{2m}+2C+2Dk^{2}\right)\left(\frac{\hbar^{2}k^{2}}{2m}+2\Delta\right)}dk\nonumber\\
		&=&\frac{1}{8\pi^{2}}\left(\frac{2m}{\hbar^{2}}\right)^{3/2}\left(2C\right)^{5/2}\left(\frac{m^{*}}{m}\right)^{2}\int_{0}^{\infty}\left(\frac{\hbar^{2}k^{2}}{2m^{*}2C}\right)^{1/2}\sqrt{\left(\frac{\hbar^{2}k^{2}}{2m^{*}2C}+1\right)\left(\frac{\hbar^{2}k^{2}}{2m^{*}2C}+\frac{m}{m^{*}}\frac{\Delta}{C}\right)}d\left(\frac{\hbar^{2}k^{2}}{2m^{*}2C}\right)\nonumber\\
		&=&\frac{1}{8\pi^{2}}\left(\frac{2m}{\hbar^{2}}\right)^{3/2}\left(2C\right)^{5/2}\left(\frac{m^{*}}{m}\right)^{2}\int_{0}^{\infty}\left(t\right)^{1/2}\sqrt{\left(t+1\right)\left(t+\frac{m}{m^{*}}\frac{\Delta}{C}\right)}dt\nonumber\\
		&=&\frac{8m^{3/2}}{15\pi^{2}\hbar^{3}}\frac{1}{\left(1+\frac{4mD}{\hbar^{2}}\right)^{2}}C^{5/2}\left[\frac{15}{4}\int_{0}^{\infty}\left(t\right)^{1/2}\sqrt{\left(t+1\right)\left[t+\left(1+\frac{4mD}{\hbar^{2}}\right)\alpha\right]}dt\right]\nonumber\\
	    &=&\frac{8m^{3/2}}{15\pi^{2}\hbar^{3}}\frac{1}{\left(1+\frac{4mD}{\hbar^{2}}\right)^{2}}C^{5/2}\Bigg[\frac{15}{4}\int_{0}^{\infty}\left(t\right)^{1/2}\Bigg(\sqrt{\left(t+1\right)\left[t+\left(1+\frac{4mD}{\hbar^{2}}\right)\alpha\right]}\nonumber\\
	    &-&\left[t+\frac{1+\left(1+\frac{4mD}{\hbar^2} \right)\alpha }{2}\right]+\frac{\left[1-\left( 1+\frac{4mD}{\hbar^2}\right)\alpha  \right]^2 }{8t}  \Bigg)dt\Bigg],
	\end{eqnarray}
where the last line represents the ultraviolet divergence term. 
For a special case, in the absence of finite-range interaction ($D=a_{\text{s}}r_{\text{eff}}^{\text{s}}C/2=0$), the integral $\int_{0}^{\infty}\left(t\right)^{1/2}\sqrt{\left(t+1\right)\left[t+\left(1+\frac{4mD}{\hbar^{2}}\right)\alpha\right]}dt$ is simplified to the form of
$\int_{0}^{\infty}\left(t\right)^{1/2}\sqrt{\left(t+1\right)\left(t+\alpha\right)}dt$. In the limit of $\alpha\rightarrow 1$, we can analytically obtain the integral by Gamma function:
\begin{eqnarray}
	\label{integralG}
\int_{0}^{\infty}\sqrt{t}\left(t+1\right)dt&=&\frac{\Gamma\left(\frac{3}{2} \right)\Gamma\left(-\frac{5}{2} \right)  }{\Gamma\left( -1\right) }\nonumber\\
&=&\frac{-\frac{4\pi}{15}}{\infty}\rightarrow0.
\end{eqnarray}

\end{widetext}

\bibliography{loopyref}

\begin{thebibliography}{72}%
\makeatletter
\providecommand \@ifxundefined [1]{%
 \@ifx{#1\undefined}
}%
\providecommand \@ifnum [1]{%
 \ifnum #1\expandafter \@firstoftwo
 \else \expandafter \@secondoftwo
 \fi
}%
\providecommand \@ifx [1]{%
 \ifx #1\expandafter \@firstoftwo
 \else \expandafter \@secondoftwo
 \fi
}%
\providecommand \natexlab [1]{#1}%
\providecommand \enquote  [1]{``#1''}%
\providecommand \bibnamefont  [1]{#1}%
\providecommand \bibfnamefont [1]{#1}%
\providecommand \citenamefont [1]{#1}%
\providecommand \href@noop [0]{\@secondoftwo}%
\providecommand \href [0]{\begingroup \@sanitize@url \@href}%
\providecommand \@href[1]{\@@startlink{#1}\@@href}%
\providecommand \@@href[1]{\endgroup#1\@@endlink}%
\providecommand \@sanitize@url [0]{\catcode `\\12\catcode `\$12\catcode
  `\&12\catcode `\#12\catcode `\^12\catcode `\_12\catcode `\%12\relax}%
\providecommand \@@startlink[1]{}%
\providecommand \@@endlink[0]{}%
\providecommand \url  [0]{\begingroup\@sanitize@url \@url }%
\providecommand \@url [1]{\endgroup\@href {#1}{\urlprefix }}%
\providecommand \urlprefix  [0]{URL }%
\providecommand \Eprint [0]{\href }%
\providecommand \doibase [0]{https://doi.org/}%
\providecommand \selectlanguage [0]{\@gobble}%
\providecommand \bibinfo  [0]{\@secondoftwo}%
\providecommand \bibfield  [0]{\@secondoftwo}%
\providecommand \translation [1]{[#1]}%
\providecommand \BibitemOpen [0]{}%
\providecommand \bibitemStop [0]{}%
\providecommand \bibitemNoStop [0]{.\EOS\space}%
\providecommand \EOS [0]{\spacefactor3000\relax}%
\providecommand \BibitemShut  [1]{\csname bibitem#1\endcsname}%
\let\auto@bib@innerbib\@empty
%</preamble>
\bibitem [{\citenamefont {Bulgac}(2002)}]{Bulgac2002}%
  \BibitemOpen
  \bibfield  {author} {\bibinfo {author} {\bibfnamefont {A.}~\bibnamefont
  {Bulgac}},\ }\bibfield  {title} {\bibinfo {title} {Dilute quantum droplets},\
  }\href {https://doi.org/10.1103/PhysRevLett.89.050402} {\bibfield  {journal}
  {\bibinfo  {journal} {Phys. Rev. Lett.}\ }\textbf {\bibinfo {volume} {89}},\
  \bibinfo {pages} {050402} (\bibinfo {year} {2002})}\BibitemShut {NoStop}%
\bibitem [{\citenamefont {Ferrier-Barbut}\ \emph {et~al.}(2016)\citenamefont
  {Ferrier-Barbut}, \citenamefont {Kadau}, \citenamefont {Schmitt},
  \citenamefont {Wenzel},\ and\ \citenamefont {Pfau}}]{Ferrier2016}%
  \BibitemOpen
  \bibfield  {author} {\bibinfo {author} {\bibfnamefont {I.}~\bibnamefont
  {Ferrier-Barbut}}, \bibinfo {author} {\bibfnamefont {H.}~\bibnamefont
  {Kadau}}, \bibinfo {author} {\bibfnamefont {M.}~\bibnamefont {Schmitt}},
  \bibinfo {author} {\bibfnamefont {M.}~\bibnamefont {Wenzel}},\ and\ \bibinfo
  {author} {\bibfnamefont {T.}~\bibnamefont {Pfau}},\ }\bibfield  {title}
  {\bibinfo {title} {Observation of quantum droplets in a strongly dipolar
  {B}ose gas},\ }\href {https://doi.org/10.1103/PhysRevLett.116.215301}
  {\bibfield  {journal} {\bibinfo  {journal} {Phys. Rev. Lett.}\ }\textbf
  {\bibinfo {volume} {116}},\ \bibinfo {pages} {215301} (\bibinfo {year}
  {2016})}\BibitemShut {NoStop}%
\bibitem [{\citenamefont {Cabrera}\ \emph {et~al.}(2018)\citenamefont
  {Cabrera}, \citenamefont {Tanzi}, \citenamefont {Sanz}, \citenamefont
  {Naylor}, \citenamefont {Thomas}, \citenamefont {Cheiney},\ and\
  \citenamefont {Tarruell}}]{Science2018}%
  \BibitemOpen
  \bibfield  {author} {\bibinfo {author} {\bibfnamefont {C.~R.}\ \bibnamefont
  {Cabrera}}, \bibinfo {author} {\bibfnamefont {L.}~\bibnamefont {Tanzi}},
  \bibinfo {author} {\bibfnamefont {J.}~\bibnamefont {Sanz}}, \bibinfo {author}
  {\bibfnamefont {B.}~\bibnamefont {Naylor}}, \bibinfo {author} {\bibfnamefont
  {P.}~\bibnamefont {Thomas}}, \bibinfo {author} {\bibfnamefont
  {P.}~\bibnamefont {Cheiney}},\ and\ \bibinfo {author} {\bibfnamefont
  {L.}~\bibnamefont {Tarruell}},\ }\bibfield  {title} {\bibinfo {title}
  {Quantum liquid droplets in a mixture of {B}ose-{E}instein condensates},\
  }\href {https://doi.org/10.1126/science.aao5686} {\bibfield  {journal}
  {\bibinfo  {journal} {Science}\ }\textbf {\bibinfo {volume} {359}},\ \bibinfo
  {pages} {301} (\bibinfo {year} {2018})}\BibitemShut {NoStop}%
\bibitem [{\citenamefont {Schmitt}\ \emph {et~al.}(2016)\citenamefont
  {Schmitt}, \citenamefont {Wenzel}, \citenamefont {B{\"o}ttcher},
  \citenamefont {Ferrier-Barbut},\ and\ \citenamefont {Pfau}}]{Schmitt2016}%
  \BibitemOpen
  \bibfield  {author} {\bibinfo {author} {\bibfnamefont {M.}~\bibnamefont
  {Schmitt}}, \bibinfo {author} {\bibfnamefont {M.}~\bibnamefont {Wenzel}},
  \bibinfo {author} {\bibfnamefont {F.}~\bibnamefont {B{\"o}ttcher}}, \bibinfo
  {author} {\bibfnamefont {I.}~\bibnamefont {Ferrier-Barbut}},\ and\ \bibinfo
  {author} {\bibfnamefont {T.}~\bibnamefont {Pfau}},\ }\bibfield  {title}
  {\bibinfo {title} {Self-bound droplets of a dilute magnetic quantum liquid},\
  }\href {https://doi.org/10.1038/nature20126} {\bibfield  {journal} {\bibinfo
  {journal} {Nature}\ }\textbf {\bibinfo {volume} {539}},\ \bibinfo {pages}
  {259} (\bibinfo {year} {2016})}\BibitemShut {NoStop}%
\bibitem [{\citenamefont {Hu}\ and\ \citenamefont
  {Liu}(2020{\natexlab{a}})}]{Hu2020}%
  \BibitemOpen
  \bibfield  {author} {\bibinfo {author} {\bibfnamefont {H.}~\bibnamefont
  {Hu}}\ and\ \bibinfo {author} {\bibfnamefont {X.-J.}\ \bibnamefont {Liu}},\
  }\bibfield  {title} {\bibinfo {title} {Consistent theory of self-bound
  quantum droplets with bosonic pairing},\ }\href
  {https://doi.org/10.1103/PhysRevLett.125.195302} {\bibfield  {journal}
  {\bibinfo  {journal} {Phys. Rev. Lett.}\ }\textbf {\bibinfo {volume} {125}},\
  \bibinfo {pages} {195302} (\bibinfo {year} {2020}{\natexlab{a}})}\BibitemShut
  {NoStop}%
\bibitem [{\citenamefont {Nagaosa}(1999)}]{Nagaosa1999}%
  \BibitemOpen
  \bibfield  {author} {\bibinfo {author} {\bibfnamefont {N.}~\bibnamefont
  {Nagaosa}},\ }\href@noop {} {\emph {\bibinfo {title} {Quantum Field Theory in
  Condensed Matter Physics}}}\ (\bibinfo  {publisher} {Springer Science \&
  Business Media},\ \bibinfo {year} {1999})\BibitemShut {NoStop}%
\bibitem [{\citenamefont {Pitaevskii}\ and\ \citenamefont
  {Stringari}(2016)}]{Pitaevskii2016}%
  \BibitemOpen
  \bibfield  {author} {\bibinfo {author} {\bibfnamefont {L.}~\bibnamefont
  {Pitaevskii}}\ and\ \bibinfo {author} {\bibfnamefont {S.}~\bibnamefont
  {Stringari}},\ }\href@noop {} {\emph {\bibinfo {title} {Bose-Einstein
  Condensation and Superfluidity}}}\ (\bibinfo  {publisher} {Oxford University
  Press},\ \bibinfo {year} {2016})\BibitemShut {NoStop}%
\bibitem [{\citenamefont {Peskin}(1995)}]{Peskin1995}%
  \BibitemOpen
  \bibfield  {author} {\bibinfo {author} {\bibfnamefont {M.~E.}\ \bibnamefont
  {Peskin}},\ }\href@noop {} {\emph {\bibinfo {title} {An Introduction to
  Quantum Field Theory}}}\ (\bibinfo  {publisher} {CRC Press},\ \bibinfo {year}
  {1995})\BibitemShut {NoStop}%
\bibitem [{\citenamefont {Petrov}(2015)}]{Petrov2015}%
  \BibitemOpen
  \bibfield  {author} {\bibinfo {author} {\bibfnamefont {D.~S.}\ \bibnamefont
  {Petrov}},\ }\bibfield  {title} {\bibinfo {title} {Quantum mechanical
  stabilization of a collapsing {B}ose-{B}ose mixture},\ }\href
  {https://doi.org/10.1103/PhysRevLett.115.155302} {\bibfield  {journal}
  {\bibinfo  {journal} {Phys. Rev. Lett.}\ }\textbf {\bibinfo {volume} {115}},\
  \bibinfo {pages} {155302} (\bibinfo {year} {2015})}\BibitemShut {NoStop}%
\bibitem [{\citenamefont {McMillan}(1965)}]{McMillan1965}%
  \BibitemOpen
  \bibfield  {author} {\bibinfo {author} {\bibfnamefont {W.~L.}\ \bibnamefont
  {McMillan}},\ }\bibfield  {title} {\bibinfo {title} {Ground state of liquid
  ${\mathrm{he}}^{4}$},\ }\href {https://doi.org/10.1103/PhysRev.138.A442}
  {\bibfield  {journal} {\bibinfo  {journal} {Phys. Rev.}\ }\textbf {\bibinfo
  {volume} {138}},\ \bibinfo {pages} {A442} (\bibinfo {year}
  {1965})}\BibitemShut {NoStop}%
\bibitem [{\citenamefont {Harms}\ \emph {et~al.}(1998)\citenamefont {Harms},
  \citenamefont {Toennies},\ and\ \citenamefont {Dalfovo}}]{Harms1998}%
  \BibitemOpen
  \bibfield  {author} {\bibinfo {author} {\bibfnamefont {J.}~\bibnamefont
  {Harms}}, \bibinfo {author} {\bibfnamefont {J.~P.}\ \bibnamefont
  {Toennies}},\ and\ \bibinfo {author} {\bibfnamefont {F.}~\bibnamefont
  {Dalfovo}},\ }\bibfield  {title} {\bibinfo {title} {Density of superfluid
  helium droplets},\ }\href {https://doi.org/10.1103/PhysRevB.58.3341}
  {\bibfield  {journal} {\bibinfo  {journal} {Phys. Rev. B}\ }\textbf {\bibinfo
  {volume} {58}},\ \bibinfo {pages} {3341} (\bibinfo {year}
  {1998})}\BibitemShut {NoStop}%
\bibitem [{\citenamefont {Dalfovo}\ and\ \citenamefont
  {Stringari}(2001)}]{Dalfovo2001}%
  \BibitemOpen
  \bibfield  {author} {\bibinfo {author} {\bibfnamefont {F.}~\bibnamefont
  {Dalfovo}}\ and\ \bibinfo {author} {\bibfnamefont {S.}~\bibnamefont
  {Stringari}},\ }\bibfield  {title} {\bibinfo {title} {Helium nanodroplets and
  trapped {B}ose–{E}instein condensates as prototypes of finite quantum
  fluids},\ }\href {https://doi.org/10.1063/1.1424926} {\bibfield  {journal}
  {\bibinfo  {journal} {J. Chem. Phys.}\ }\textbf {\bibinfo {volume} {115}},\
  \bibinfo {pages} {10078} (\bibinfo {year} {2001})}\BibitemShut {NoStop}%
\bibitem [{\citenamefont {Ancilotto}\ \emph {et~al.}(2017)\citenamefont
  {Ancilotto}, \citenamefont {Barranco}, \citenamefont {Coppens}, \citenamefont
  {Eloranta}, \citenamefont {Halberstadt}, \citenamefont {Hernando},
  \citenamefont {Mateo},\ and\ \citenamefont {Pi}}]{Francesco2017}%
  \BibitemOpen
  \bibfield  {author} {\bibinfo {author} {\bibfnamefont {F.}~\bibnamefont
  {Ancilotto}}, \bibinfo {author} {\bibfnamefont {M.}~\bibnamefont {Barranco}},
  \bibinfo {author} {\bibfnamefont {F.}~\bibnamefont {Coppens}}, \bibinfo
  {author} {\bibfnamefont {J.}~\bibnamefont {Eloranta}}, \bibinfo {author}
  {\bibfnamefont {N.}~\bibnamefont {Halberstadt}}, \bibinfo {author}
  {\bibfnamefont {A.}~\bibnamefont {Hernando}}, \bibinfo {author}
  {\bibfnamefont {D.}~\bibnamefont {Mateo}},\ and\ \bibinfo {author}
  {\bibfnamefont {M.}~\bibnamefont {Pi}},\ }\bibfield  {title} {\bibinfo
  {title} {Density functional theory of doped superfluid liquid helium and
  nanodroplets},\ }\href {https://doi.org/10.1080/0144235X.2017.1351672}
  {\bibfield  {journal} {\bibinfo  {journal} {Int. Rev. Phys. Chem.}\ }\textbf
  {\bibinfo {volume} {36}},\ \bibinfo {pages} {621} (\bibinfo {year}
  {2017})}\BibitemShut {NoStop}%
\bibitem [{\citenamefont {Pang}(2020)}]{Pang2020}%
  \BibitemOpen
  \bibfield  {author} {\bibinfo {author} {\bibfnamefont {T.}~\bibnamefont
  {Pang}},\ }\bibfield  {title} {\bibinfo {title} {The breakup of a helium
  cluster after removing attractive interaction among a significant number of
  atoms in the cluster},\ }\href {https://doi.org/10.1038/s41598-020-62732-2}
  {\bibfield  {journal} {\bibinfo  {journal} {Sci. Rep.}\ }\textbf {\bibinfo
  {volume} {10}},\ \bibinfo {pages} {5767} (\bibinfo {year}
  {2020})}\BibitemShut {NoStop}%
\bibitem [{\citenamefont {Bloch}\ \emph {et~al.}(2008)\citenamefont {Bloch},
  \citenamefont {Dalibard},\ and\ \citenamefont {Zwerger}}]{Bloch2008}%
  \BibitemOpen
  \bibfield  {author} {\bibinfo {author} {\bibfnamefont {I.}~\bibnamefont
  {Bloch}}, \bibinfo {author} {\bibfnamefont {J.}~\bibnamefont {Dalibard}},\
  and\ \bibinfo {author} {\bibfnamefont {W.}~\bibnamefont {Zwerger}},\
  }\bibfield  {title} {\bibinfo {title} {Many-body physics with ultracold
  gases},\ }\href {https://doi.org/10.1103/RevModPhys.80.885} {\bibfield
  {journal} {\bibinfo  {journal} {Rev. Mod. Phys.}\ }\textbf {\bibinfo {volume}
  {80}},\ \bibinfo {pages} {885} (\bibinfo {year} {2008})}\BibitemShut
  {NoStop}%
\bibitem [{\citenamefont {Boudjem\^aa}(2018)}]{Boudjemaa2018}%
  \BibitemOpen
  \bibfield  {author} {\bibinfo {author} {\bibfnamefont {A.}~\bibnamefont
  {Boudjem\^aa}},\ }\bibfield  {title} {\bibinfo {title} {{Fluctuations and
  quantum self-bound droplets in a dipolar Bose-Bose mixture}},\ }\href
  {https://doi.org/10.1103/PhysRevA.98.033612} {\bibfield  {journal} {\bibinfo
  {journal} {Phys. Rev. A}\ }\textbf {\bibinfo {volume} {98}},\ \bibinfo
  {pages} {033612} (\bibinfo {year} {2018})}\BibitemShut {NoStop}%
\bibitem [{\citenamefont {Abbas}\ and\ \citenamefont
  {Boudjem\^aa}(2023)}]{Abbas2023}%
  \BibitemOpen
  \bibfield  {author} {\bibinfo {author} {\bibfnamefont {K.}~\bibnamefont
  {Abbas}}\ and\ \bibinfo {author} {\bibfnamefont {A.}~\bibnamefont
  {Boudjem\^aa}},\ }\bibfield  {title} {\bibinfo {title} {{Quantum liquid
  droplets in Bose mixtures with weak disorder}},\ }\href
  {https://doi.org/10.1103/PhysRevA.107.033306} {\bibfield  {journal} {\bibinfo
   {journal} {Phys. Rev. A}\ }\textbf {\bibinfo {volume} {107}},\ \bibinfo
  {pages} {033306} (\bibinfo {year} {2023})}\BibitemShut {NoStop}%
\bibitem [{\citenamefont {Petrov}\ and\ \citenamefont
  {Astrakharchik}(2016)}]{Petrov2016}%
  \BibitemOpen
  \bibfield  {author} {\bibinfo {author} {\bibfnamefont {D.~S.}\ \bibnamefont
  {Petrov}}\ and\ \bibinfo {author} {\bibfnamefont {G.~E.}\ \bibnamefont
  {Astrakharchik}},\ }\bibfield  {title} {\bibinfo {title} {{Ultradilute
  low-dimensional liquids}},\ }\href
  {https://doi.org/10.1103/PhysRevLett.117.100401} {\bibfield  {journal}
  {\bibinfo  {journal} {Phys. Rev. Lett.}\ }\textbf {\bibinfo {volume} {117}},\
  \bibinfo {pages} {100401} (\bibinfo {year} {2016})}\BibitemShut {NoStop}%
\bibitem [{\citenamefont {He}\ \emph {et~al.}(2023)\citenamefont {He},
  \citenamefont {Li}, \citenamefont {Yi},\ and\ \citenamefont {Yu}}]{HeLi2023}%
  \BibitemOpen
  \bibfield  {author} {\bibinfo {author} {\bibfnamefont {L.}~\bibnamefont
  {He}}, \bibinfo {author} {\bibfnamefont {H.}~\bibnamefont {Li}}, \bibinfo
  {author} {\bibfnamefont {W.}~\bibnamefont {Yi}},\ and\ \bibinfo {author}
  {\bibfnamefont {Z.-Q.}\ \bibnamefont {Yu}},\ }\bibfield  {title} {\bibinfo
  {title} {Quantum criticality of liquid-gas transition in a binary {B}ose
  mixture},\ }\href {https://doi.org/10.1103/PhysRevLett.130.193001} {\bibfield
   {journal} {\bibinfo  {journal} {Phys. Rev. Lett.}\ }\textbf {\bibinfo
  {volume} {130}},\ \bibinfo {pages} {193001} (\bibinfo {year}
  {2023})}\BibitemShut {NoStop}%
\bibitem [{\citenamefont {Dalfovo}\ \emph {et~al.}(1999)\citenamefont
  {Dalfovo}, \citenamefont {Giorgini}, \citenamefont {Pitaevskii},\ and\
  \citenamefont {Stringari}}]{Dalfovo1999}%
  \BibitemOpen
  \bibfield  {author} {\bibinfo {author} {\bibfnamefont {F.}~\bibnamefont
  {Dalfovo}}, \bibinfo {author} {\bibfnamefont {S.}~\bibnamefont {Giorgini}},
  \bibinfo {author} {\bibfnamefont {L.~P.}\ \bibnamefont {Pitaevskii}},\ and\
  \bibinfo {author} {\bibfnamefont {S.}~\bibnamefont {Stringari}},\ }\bibfield
  {title} {\bibinfo {title} {Theory of {B}ose-{E}instein condensation in
  trapped gases},\ }\href {https://doi.org/10.1103/RevModPhys.71.463}
  {\bibfield  {journal} {\bibinfo  {journal} {Rev. Mod. Phys.}\ }\textbf
  {\bibinfo {volume} {71}},\ \bibinfo {pages} {463} (\bibinfo {year}
  {1999})}\BibitemShut {NoStop}%
\bibitem [{\citenamefont {Salasnich}(2017)}]{Salasnich2017}%
  \BibitemOpen
  \bibfield  {author} {\bibinfo {author} {\bibfnamefont {L.}~\bibnamefont
  {Salasnich}},\ }\bibfield  {title} {\bibinfo {title} {{Nonuniversal equation
  of state of the two-dimensional Bose gas}},\ }\href
  {https://doi.org/10.1103/PhysRevLett.118.130402} {\bibfield  {journal}
  {\bibinfo  {journal} {Phys. Rev. Lett.}\ }\textbf {\bibinfo {volume} {118}},\
  \bibinfo {pages} {130402} (\bibinfo {year} {2017})}\BibitemShut {NoStop}%
\bibitem [{\citenamefont {Cappellaro}\ and\ \citenamefont
  {Salasnich}(2017)}]{Cappe2017}%
  \BibitemOpen
  \bibfield  {author} {\bibinfo {author} {\bibfnamefont {A.}~\bibnamefont
  {Cappellaro}}\ and\ \bibinfo {author} {\bibfnamefont {L.}~\bibnamefont
  {Salasnich}},\ }\bibfield  {title} {\bibinfo {title} {{Thermal field theory
  of bosonic gases with finite-range effective interaction}},\ }\href
  {https://doi.org/10.1103/PhysRevA.95.033627} {\bibfield  {journal} {\bibinfo
  {journal} {Phys. Rev. A}\ }\textbf {\bibinfo {volume} {95}},\ \bibinfo
  {pages} {033627} (\bibinfo {year} {2017})}\BibitemShut {NoStop}%
\bibitem [{\citenamefont {Cappellaro}\ \emph {et~al.}(2017)\citenamefont
  {Cappellaro}, \citenamefont {Macr{\`\i}}, \citenamefont {Bertacco},\ and\
  \citenamefont {Salasnich}}]{Cappellaro2017}%
  \BibitemOpen
  \bibfield  {author} {\bibinfo {author} {\bibfnamefont {A.}~\bibnamefont
  {Cappellaro}}, \bibinfo {author} {\bibfnamefont {T.}~\bibnamefont
  {Macr{\`\i}}}, \bibinfo {author} {\bibfnamefont {G.~F.}\ \bibnamefont
  {Bertacco}},\ and\ \bibinfo {author} {\bibfnamefont {L.}~\bibnamefont
  {Salasnich}},\ }\bibfield  {title} {\bibinfo {title} {Equation of state and
  self-bound droplet in {R}abi-coupled {B}ose mixtures},\ }\href
  {https://doi.org/10.1038/s41598-017-13647-y} {\bibfield  {journal} {\bibinfo
  {journal} {Sci. Rep.}\ }\textbf {\bibinfo {volume} {7}},\ \bibinfo {pages}
  {13358} (\bibinfo {year} {2017})}\BibitemShut {NoStop}%
\bibitem [{\citenamefont {Tononi}\ \emph {et~al.}(2018)\citenamefont {Tononi},
  \citenamefont {Cappellaro},\ and\ \citenamefont {Salasnich}}]{Tononi2018}%
  \BibitemOpen
  \bibfield  {author} {\bibinfo {author} {\bibfnamefont {A.}~\bibnamefont
  {Tononi}}, \bibinfo {author} {\bibfnamefont {A.}~\bibnamefont {Cappellaro}},\
  and\ \bibinfo {author} {\bibfnamefont {L.}~\bibnamefont {Salasnich}},\
  }\bibfield  {title} {\bibinfo {title} {{Condensation and superfluidity of
  dilute Bose gases with finite-range interaction}},\ }\href
  {https://doi.org/10.1088/1367-2630/aaf75e} {\bibfield  {journal} {\bibinfo
  {journal} {New J. Phys.}\ }\textbf {\bibinfo {volume} {20}},\ \bibinfo
  {pages} {125007} (\bibinfo {year} {2018})}\BibitemShut {NoStop}%
\bibitem [{\citenamefont {Yin}\ \emph {et~al.}(2020)\citenamefont {Yin},
  \citenamefont {Hu},\ and\ \citenamefont {Liu}}]{Yin2020}%
  \BibitemOpen
  \bibfield  {author} {\bibinfo {author} {\bibfnamefont {X.~Y.}\ \bibnamefont
  {Yin}}, \bibinfo {author} {\bibfnamefont {H.}~\bibnamefont {Hu}},\ and\
  \bibinfo {author} {\bibfnamefont {X.-J.}\ \bibnamefont {Liu}},\ }\bibfield
  {title} {\bibinfo {title} {Few-body perspective of a quantum anomaly in
  two-dimensional {F}ermi gases},\ }\href
  {https://doi.org/10.1103/PhysRevLett.124.013401} {\bibfield  {journal}
  {\bibinfo  {journal} {Phys. Rev. Lett.}\ }\textbf {\bibinfo {volume} {124}},\
  \bibinfo {pages} {013401} (\bibinfo {year} {2020})}\BibitemShut {NoStop}%
\bibitem [{\citenamefont {Yu}\ \emph {et~al.}(2024)\citenamefont {Yu},
  \citenamefont {Ye},\ and\ \citenamefont {Liang}}]{Yu2024}%
  \BibitemOpen
  \bibfield  {author} {\bibinfo {author} {\bibfnamefont {T.}~\bibnamefont
  {Yu}}, \bibinfo {author} {\bibfnamefont {X.}~\bibnamefont {Ye}},\ and\
  \bibinfo {author} {\bibfnamefont {Z.}~\bibnamefont {Liang}},\ }\bibfield
  {title} {\bibinfo {title} {Interaction-induced dimensional crossover from
  fully three-dimensional to one-dimensional spaces},\ }\href
  {https://doi.org/10.1103/PhysRevA.110.013304} {\bibfield  {journal} {\bibinfo
   {journal} {Phys. Rev. A}\ }\textbf {\bibinfo {volume} {110}},\ \bibinfo
  {pages} {013304} (\bibinfo {year} {2024})}\BibitemShut {NoStop}%
\bibitem [{\citenamefont {Ye}\ \emph {et~al.}(2024)\citenamefont {Ye},
  \citenamefont {Yu},\ and\ \citenamefont {Liang}}]{Ye2024}%
  \BibitemOpen
  \bibfield  {author} {\bibinfo {author} {\bibfnamefont {X.}~\bibnamefont
  {Ye}}, \bibinfo {author} {\bibfnamefont {T.}~\bibnamefont {Yu}},\ and\
  \bibinfo {author} {\bibfnamefont {Z.}~\bibnamefont {Liang}},\ }\bibfield
  {title} {\bibinfo {title} {{Nonuniversal equation of state of a
  quasi-two-dimensional Bose gas in dimensional crossover}},\ }\href
  {https://doi.org/10.1103/PhysRevA.109.063304} {\bibfield  {journal} {\bibinfo
   {journal} {Phys. Rev. A}\ }\textbf {\bibinfo {volume} {109}},\ \bibinfo
  {pages} {063304} (\bibinfo {year} {2024})}\BibitemShut {NoStop}%
\bibitem [{\citenamefont {Zhang}\ and\ \citenamefont
  {Liang}(2024)}]{Zhang2024}%
  \BibitemOpen
  \bibfield  {author} {\bibinfo {author} {\bibfnamefont {Y.}~\bibnamefont
  {Zhang}}\ and\ \bibinfo {author} {\bibfnamefont {Z.}~\bibnamefont {Liang}},\
  }\bibfield  {title} {\bibinfo {title} {{Cornwall-Jackiw-Tomboulis effective
  field theory and the nonuniversal equation of state of an ultracold Bose
  gas}},\ }\href {https://doi.org/10.1103/PhysRevA.110.043318} {\bibfield
  {journal} {\bibinfo  {journal} {Phys. Rev. A}\ }\textbf {\bibinfo {volume}
  {110}},\ \bibinfo {pages} {043318} (\bibinfo {year} {2024})}\BibitemShut
  {NoStop}%
\bibitem [{\citenamefont {Chiquillo}(2025)}]{Emerson2025}%
  \BibitemOpen
  \bibfield  {author} {\bibinfo {author} {\bibfnamefont {E.}~\bibnamefont
  {Chiquillo}},\ }\bibfield  {title} {\bibinfo {title} {Nonuniversal equation
  of state for {R}abi-coupled bosonic gases: A droplet phase},\ }\href
  {https://doi.org/https://doi.org/10.1016/j.aop.2025.170071} {\bibfield
  {journal} {\bibinfo  {journal} {Ann. Phys. (NY)}\ }\textbf {\bibinfo {volume}
  {479}},\ \bibinfo {pages} {170071} (\bibinfo {year} {2025})}\BibitemShut
  {NoStop}%
\bibitem [{\citenamefont {Petrov}(2018)}]{Petrov2018}%
  \BibitemOpen
  \bibfield  {author} {\bibinfo {author} {\bibfnamefont {D.~S.}\ \bibnamefont
  {Petrov}},\ }\bibfield  {title} {\bibinfo {title} {Liquid beyond the van der
  {W}aals paradigm},\ }\href {https://doi.org/10.1038/s41567-018-0052-9}
  {\bibfield  {journal} {\bibinfo  {journal} {Nat. Phys.}\ }\textbf {\bibinfo
  {volume} {14}},\ \bibinfo {pages} {211} (\bibinfo {year} {2018})}\BibitemShut
  {NoStop}%
\bibitem [{\citenamefont {Chiquillo}(2019)}]{Chiquillo2019}%
  \BibitemOpen
  \bibfield  {author} {\bibinfo {author} {\bibfnamefont {E.}~\bibnamefont
  {Chiquillo}},\ }\bibfield  {title} {\bibinfo {title} {Low-dimensional
  self-bound quantum {R}abi-coupled bosonic droplets},\ }\href
  {https://doi.org/10.1103/PhysRevA.99.051601} {\bibfield  {journal} {\bibinfo
  {journal} {Phys. Rev. A}\ }\textbf {\bibinfo {volume} {99}},\ \bibinfo
  {pages} {051601} (\bibinfo {year} {2019})}\BibitemShut {NoStop}%
\bibitem [{\citenamefont {Zhang}\ and\ \citenamefont {Yin}(2025)}]{Zhang2025}%
  \BibitemOpen
  \bibfield  {author} {\bibinfo {author} {\bibfnamefont {F.}~\bibnamefont
  {Zhang}}\ and\ \bibinfo {author} {\bibfnamefont {L.}~\bibnamefont {Yin}},\
  }\bibfield  {title} {\bibinfo {title} {Density-functional theory of quantum
  droplets},\ }\href {https://doi.org/10.1088/0256-307X/42/1/010302} {\bibfield
   {journal} {\bibinfo  {journal} {Chin. Phys. Lett.}\ }\textbf {\bibinfo
  {volume} {42}},\ \bibinfo {pages} {010302} (\bibinfo {year}
  {2025})}\BibitemShut {NoStop}%
\bibitem [{\citenamefont {Flachi}\ and\ \citenamefont
  {Tanaka}(2025)}]{Flachi2025}%
  \BibitemOpen
  \bibfield  {author} {\bibinfo {author} {\bibfnamefont {A.}~\bibnamefont
  {Flachi}}\ and\ \bibinfo {author} {\bibfnamefont {T.}~\bibnamefont
  {Tanaka}},\ }\bibfield  {title} {\bibinfo {title} {Quantum droplets in curved
  space},\ }\href {https://doi.org/10.1103/145z-6dnb} {\bibfield  {journal}
  {\bibinfo  {journal} {Phys. Rev. Lett.}\ }\textbf {\bibinfo {volume} {135}},\
  \bibinfo {pages} {231601} (\bibinfo {year} {2025})}\BibitemShut {NoStop}%
\bibitem [{\citenamefont {Englezos}\ \emph {et~al.}(2025)\citenamefont
  {Englezos}, \citenamefont {Schmelcher},\ and\ \citenamefont
  {Mistakidis}}]{Englezos2025}%
  \BibitemOpen
  \bibfield  {author} {\bibinfo {author} {\bibfnamefont {I.~A.}\ \bibnamefont
  {Englezos}}, \bibinfo {author} {\bibfnamefont {P.}~\bibnamefont
  {Schmelcher}},\ and\ \bibinfo {author} {\bibfnamefont {S.~I.}\ \bibnamefont
  {Mistakidis}},\ }\bibfield  {title} {\bibinfo {title} {{Multicomponent
  one-dimensional quantum droplets across the mean-field stability regime}},\
  }\href {https://doi.org/10.21468/SciPostPhys.19.5.133} {\bibfield  {journal}
  {\bibinfo  {journal} {SciPost Phys.}\ }\textbf {\bibinfo {volume} {19}},\
  \bibinfo {pages} {133} (\bibinfo {year} {2025})}\BibitemShut {NoStop}%
\bibitem [{\citenamefont {Ferioli}\ \emph {et~al.}(2020)\citenamefont
  {Ferioli}, \citenamefont {Semeghini}, \citenamefont {Terradas-Brians\'o},
  \citenamefont {Masi}, \citenamefont {Fattori},\ and\ \citenamefont
  {Modugno}}]{Ferioli2020}%
  \BibitemOpen
  \bibfield  {author} {\bibinfo {author} {\bibfnamefont {G.}~\bibnamefont
  {Ferioli}}, \bibinfo {author} {\bibfnamefont {G.}~\bibnamefont {Semeghini}},
  \bibinfo {author} {\bibfnamefont {S.}~\bibnamefont {Terradas-Brians\'o}},
  \bibinfo {author} {\bibfnamefont {L.}~\bibnamefont {Masi}}, \bibinfo {author}
  {\bibfnamefont {M.}~\bibnamefont {Fattori}},\ and\ \bibinfo {author}
  {\bibfnamefont {M.}~\bibnamefont {Modugno}},\ }\bibfield  {title} {\bibinfo
  {title} {Dynamical formation of quantum droplets in a $^{39}\mathrm{K}$
  mixture},\ }\href {https://doi.org/10.1103/PhysRevResearch.2.013269}
  {\bibfield  {journal} {\bibinfo  {journal} {Phys. Rev. Res.}\ }\textbf
  {\bibinfo {volume} {2}},\ \bibinfo {pages} {013269} (\bibinfo {year}
  {2020})}\BibitemShut {NoStop}%
\bibitem [{\citenamefont {Banerjee}\ \emph {et~al.}(2025)\citenamefont
  {Banerjee}, \citenamefont {Zhou}, \citenamefont {Tiwari}, \citenamefont
  {Tamura}, \citenamefont {Li}, \citenamefont {Kevrekidis}, \citenamefont
  {Mistakidis}, \citenamefont {Walther},\ and\ \citenamefont
  {Hung}}]{Banerjee2025}%
  \BibitemOpen
  \bibfield  {author} {\bibinfo {author} {\bibfnamefont {S.}~\bibnamefont
  {Banerjee}}, \bibinfo {author} {\bibfnamefont {K.}~\bibnamefont {Zhou}},
  \bibinfo {author} {\bibfnamefont {S.~K.}\ \bibnamefont {Tiwari}}, \bibinfo
  {author} {\bibfnamefont {H.}~\bibnamefont {Tamura}}, \bibinfo {author}
  {\bibfnamefont {R.}~\bibnamefont {Li}}, \bibinfo {author} {\bibfnamefont
  {P.}~\bibnamefont {Kevrekidis}}, \bibinfo {author} {\bibfnamefont {S.~I.}\
  \bibnamefont {Mistakidis}}, \bibinfo {author} {\bibfnamefont
  {V.}~\bibnamefont {Walther}},\ and\ \bibinfo {author} {\bibfnamefont {C.-L.}\
  \bibnamefont {Hung}},\ }\bibfield  {title} {\bibinfo {title} {Collapse of a
  quantum vortex in an attractive two-dimensional {B}ose gas},\ }\href
  {https://doi.org/10.1103/c6wx-zc9x} {\bibfield  {journal} {\bibinfo
  {journal} {Phys. Rev. Lett.}\ }\textbf {\bibinfo {volume} {135}},\ \bibinfo
  {pages} {073401} (\bibinfo {year} {2025})}\BibitemShut {NoStop}%
\bibitem [{\citenamefont {Cavicchioli}\ \emph {et~al.}(2025)\citenamefont
  {Cavicchioli}, \citenamefont {Fort}, \citenamefont {Ancilotto}, \citenamefont
  {Modugno}, \citenamefont {Minardi},\ and\ \citenamefont
  {Burchianti}}]{Cavicchioli2025}%
  \BibitemOpen
  \bibfield  {author} {\bibinfo {author} {\bibfnamefont {L.}~\bibnamefont
  {Cavicchioli}}, \bibinfo {author} {\bibfnamefont {C.}~\bibnamefont {Fort}},
  \bibinfo {author} {\bibfnamefont {F.}~\bibnamefont {Ancilotto}}, \bibinfo
  {author} {\bibfnamefont {M.}~\bibnamefont {Modugno}}, \bibinfo {author}
  {\bibfnamefont {F.}~\bibnamefont {Minardi}},\ and\ \bibinfo {author}
  {\bibfnamefont {A.}~\bibnamefont {Burchianti}},\ }\bibfield  {title}
  {\bibinfo {title} {Dynamical formation of multiple quantum droplets in a
  {B}ose-{B}ose mixture},\ }\href
  {https://doi.org/10.1103/PhysRevLett.134.093401} {\bibfield  {journal}
  {\bibinfo  {journal} {Phys. Rev. Lett.}\ }\textbf {\bibinfo {volume} {134}},\
  \bibinfo {pages} {093401} (\bibinfo {year} {2025})}\BibitemShut {NoStop}%
\bibitem [{\citenamefont {Ye}\ and\ \citenamefont {Liang}(2025)}]{Xiaoran2025}%
  \BibitemOpen
  \bibfield  {author} {\bibinfo {author} {\bibfnamefont {X.}~\bibnamefont
  {Ye}}\ and\ \bibinfo {author} {\bibfnamefont {Z.}~\bibnamefont {Liang}},\
  }\bibfield  {title} {\bibinfo {title} {Probing $p$-wave effects in
  spin-density separation of {B}ose mixtures with the dynamic structure
  factor},\ }\href {https://doi.org/10.1103/PhysRevA.111.043311} {\bibfield
  {journal} {\bibinfo  {journal} {Phys. Rev. A}\ }\textbf {\bibinfo {volume}
  {111}},\ \bibinfo {pages} {043311} (\bibinfo {year} {2025})}\BibitemShut
  {NoStop}%
\bibitem [{\citenamefont {Lee}\ \emph {et~al.}(1957)\citenamefont {Lee},
  \citenamefont {Huang},\ and\ \citenamefont {Yang}}]{Lee1957}%
  \BibitemOpen
  \bibfield  {author} {\bibinfo {author} {\bibfnamefont {T.~D.}\ \bibnamefont
  {Lee}}, \bibinfo {author} {\bibfnamefont {K.}~\bibnamefont {Huang}},\ and\
  \bibinfo {author} {\bibfnamefont {C.~N.}\ \bibnamefont {Yang}},\ }\bibfield
  {title} {\bibinfo {title} {Eigenvalues and eigenfunctions of a {B}ose system
  of hard spheres and its low-temperature properties},\ }\href
  {https://doi.org/10.1103/PhysRev.106.1135} {\bibfield  {journal} {\bibinfo
  {journal} {Phys. Rev.}\ }\textbf {\bibinfo {volume} {106}},\ \bibinfo {pages}
  {1135} (\bibinfo {year} {1957})}\BibitemShut {NoStop}%
\bibitem [{\citenamefont {Lee}\ and\ \citenamefont {Yang}(1957)}]{Yang1957}%
  \BibitemOpen
  \bibfield  {author} {\bibinfo {author} {\bibfnamefont {T.~D.}\ \bibnamefont
  {Lee}}\ and\ \bibinfo {author} {\bibfnamefont {C.~N.}\ \bibnamefont {Yang}},\
  }\bibfield  {title} {\bibinfo {title} {Many-body problem in quantum mechanics
  and quantum statistical mechanics},\ }\href
  {https://doi.org/10.1103/PhysRev.105.1119} {\bibfield  {journal} {\bibinfo
  {journal} {Phys. Rev.}\ }\textbf {\bibinfo {volume} {105}},\ \bibinfo {pages}
  {1119} (\bibinfo {year} {1957})}\BibitemShut {NoStop}%
\bibitem [{\citenamefont {Hu}\ \emph {et~al.}(2020)\citenamefont {Hu},
  \citenamefont {Wang},\ and\ \citenamefont {Liu}}]{Hu2020b}%
  \BibitemOpen
  \bibfield  {author} {\bibinfo {author} {\bibfnamefont {H.}~\bibnamefont
  {Hu}}, \bibinfo {author} {\bibfnamefont {J.}~\bibnamefont {Wang}},\ and\
  \bibinfo {author} {\bibfnamefont {X.-J.}\ \bibnamefont {Liu}},\ }\bibfield
  {title} {\bibinfo {title} {Microscopic pairing theory of a binary {B}ose
  mixture with interspecies attractions: Bosonic {BEC-BCS} crossover and
  ultradilute low-dimensional quantum droplets},\ }\href
  {https://doi.org/10.1103/PhysRevA.102.043301} {\bibfield  {journal} {\bibinfo
   {journal} {Phys. Rev. A}\ }\textbf {\bibinfo {volume} {102}},\ \bibinfo
  {pages} {043301} (\bibinfo {year} {2020})}\BibitemShut {NoStop}%
\bibitem [{\citenamefont {Hu}\ and\ \citenamefont
  {Liu}(2020{\natexlab{b}})}]{Hu2020d}%
  \BibitemOpen
  \bibfield  {author} {\bibinfo {author} {\bibfnamefont {H.}~\bibnamefont
  {Hu}}\ and\ \bibinfo {author} {\bibfnamefont {X.-J.}\ \bibnamefont {Liu}},\
  }\bibfield  {title} {\bibinfo {title} {Microscopic derivation of the extended
  {G}ross-{P}itaevskii equation for quantum droplets in binary {B}ose
  mixtures},\ }\href {https://doi.org/10.1103/PhysRevA.102.043302} {\bibfield
  {journal} {\bibinfo  {journal} {Phys. Rev. A}\ }\textbf {\bibinfo {volume}
  {102}},\ \bibinfo {pages} {043302} (\bibinfo {year}
  {2020}{\natexlab{b}})}\BibitemShut {NoStop}%
\bibitem [{\citenamefont {Hu}\ \emph {et~al.}(2025)\citenamefont {Hu},
  \citenamefont {Wang}, \citenamefont {Pu},\ and\ \citenamefont
  {Liu}}]{Hu2025}%
  \BibitemOpen
  \bibfield  {author} {\bibinfo {author} {\bibfnamefont {H.}~\bibnamefont
  {Hu}}, \bibinfo {author} {\bibfnamefont {J.}~\bibnamefont {Wang}}, \bibinfo
  {author} {\bibfnamefont {H.}~\bibnamefont {Pu}},\ and\ \bibinfo {author}
  {\bibfnamefont {X.-J.}\ \bibnamefont {Liu}},\ }\bibfield  {title} {\bibinfo
  {title} {Breakdown of the single-mode description of ultradilute quantum
  droplets in binary {B}ose mixtures: A perspective from a microscopic bosonic
  pairing theory},\ }\href {https://doi.org/10.1103/PhysRevA.111.023309}
  {\bibfield  {journal} {\bibinfo  {journal} {Phys. Rev. A}\ }\textbf {\bibinfo
  {volume} {111}},\ \bibinfo {pages} {023309} (\bibinfo {year}
  {2025})}\BibitemShut {NoStop}%
\bibitem [{\citenamefont {Bardeen}\ \emph {et~al.}(1957)\citenamefont
  {Bardeen}, \citenamefont {Cooper},\ and\ \citenamefont
  {Schrieffer}}]{Bardeen1957}%
  \BibitemOpen
  \bibfield  {author} {\bibinfo {author} {\bibfnamefont {J.}~\bibnamefont
  {Bardeen}}, \bibinfo {author} {\bibfnamefont {L.~N.}\ \bibnamefont
  {Cooper}},\ and\ \bibinfo {author} {\bibfnamefont {J.~R.}\ \bibnamefont
  {Schrieffer}},\ }\bibfield  {title} {\bibinfo {title} {Microscopic theory of
  superconductivity},\ }\href {https://doi.org/10.1103/PhysRev.106.162}
  {\bibfield  {journal} {\bibinfo  {journal} {Phys. Rev.}\ }\textbf {\bibinfo
  {volume} {106}},\ \bibinfo {pages} {162} (\bibinfo {year}
  {1957})}\BibitemShut {NoStop}%
\bibitem [{\citenamefont {Hu}\ \emph {et~al.}(2006)\citenamefont {Hu},
  \citenamefont {Liu},\ and\ \citenamefont {Drummond}}]{Hu2006}%
  \BibitemOpen
  \bibfield  {author} {\bibinfo {author} {\bibfnamefont {H.}~\bibnamefont
  {Hu}}, \bibinfo {author} {\bibfnamefont {X.-J.}\ \bibnamefont {Liu}},\ and\
  \bibinfo {author} {\bibfnamefont {P.~D.}\ \bibnamefont {Drummond}},\
  }\bibfield  {title} {\bibinfo {title} {Equation of state of a superfluid
  {F}ermi gas in the {BCS-BEC} crossover},\ }\href
  {https://doi.org/10.1209/epl/i2006-10023-y} {\bibfield  {journal} {\bibinfo
  {journal} {Europhys. Lett.}\ }\textbf {\bibinfo {volume} {74}},\ \bibinfo
  {pages} {574} (\bibinfo {year} {2006})}\BibitemShut {NoStop}%
\bibitem [{\citenamefont {He}\ \emph {et~al.}(2015)\citenamefont {He},
  \citenamefont {L\"u}, \citenamefont {Cao}, \citenamefont {Hu},\ and\
  \citenamefont {Liu}}]{He2015}%
  \BibitemOpen
  \bibfield  {author} {\bibinfo {author} {\bibfnamefont {L.}~\bibnamefont
  {He}}, \bibinfo {author} {\bibfnamefont {H.}~\bibnamefont {L\"u}}, \bibinfo
  {author} {\bibfnamefont {G.}~\bibnamefont {Cao}}, \bibinfo {author}
  {\bibfnamefont {H.}~\bibnamefont {Hu}},\ and\ \bibinfo {author}
  {\bibfnamefont {X.-J.}\ \bibnamefont {Liu}},\ }\bibfield  {title} {\bibinfo
  {title} {Quantum fluctuations in the {BCS-BEC} crossover of two-dimensional
  {F}ermi gases},\ }\href {https://doi.org/10.1103/PhysRevA.92.023620}
  {\bibfield  {journal} {\bibinfo  {journal} {Phys. Rev. A}\ }\textbf {\bibinfo
  {volume} {92}},\ \bibinfo {pages} {023620} (\bibinfo {year}
  {2015})}\BibitemShut {NoStop}%
\bibitem [{\citenamefont {Shen}\ and\ \citenamefont {Yin}(2025)}]{Shen2025}%
  \BibitemOpen
  \bibfield  {author} {\bibinfo {author} {\bibfnamefont {Z.}~\bibnamefont
  {Shen}}\ and\ \bibinfo {author} {\bibfnamefont {L.}~\bibnamefont {Yin}},\
  }\bibfield  {title} {\bibinfo {title} {Pairing transitions in a binary {B}ose
  gas},\ }\href {https://doi.org/10.1088/1674-1056/ade075} {\bibfield
  {journal} {\bibinfo  {journal} {Chin. Phys. B}\ }\textbf {\bibinfo {volume}
  {34}},\ \bibinfo {pages} {106702} (\bibinfo {year} {2025})}\BibitemShut
  {NoStop}%
\bibitem [{\citenamefont {Cikojevi\ifmmode~\acute{c}\else \'{c}\fi{}}\ \emph
  {et~al.}(2019)\citenamefont {Cikojevi\ifmmode~\acute{c}\else \'{c}\fi{}},
  \citenamefont {Marki\ifmmode~\acute{c}\else \'{c}\fi{}}, \citenamefont
  {Astrakharchik},\ and\ \citenamefont {Boronat}}]{Cikojevi2019}%
  \BibitemOpen
  \bibfield  {author} {\bibinfo {author} {\bibfnamefont {V.}~\bibnamefont
  {Cikojevi\ifmmode~\acute{c}\else \'{c}\fi{}}}, \bibinfo {author}
  {\bibfnamefont {L.~V. c.~v.}\ \bibnamefont {Marki\ifmmode~\acute{c}\else
  \'{c}\fi{}}}, \bibinfo {author} {\bibfnamefont {G.~E.}\ \bibnamefont
  {Astrakharchik}},\ and\ \bibinfo {author} {\bibfnamefont {J.}~\bibnamefont
  {Boronat}},\ }\bibfield  {title} {\bibinfo {title} {Universality in
  ultradilute liquid {B}ose-{B}ose mixtures},\ }\href
  {https://doi.org/10.1103/PhysRevA.99.023618} {\bibfield  {journal} {\bibinfo
  {journal} {Phys. Rev. A}\ }\textbf {\bibinfo {volume} {99}},\ \bibinfo
  {pages} {023618} (\bibinfo {year} {2019})}\BibitemShut {NoStop}%
\bibitem [{\citenamefont {Cikojević}\ \emph {et~al.}(2020)\citenamefont
  {Cikojević}, \citenamefont {Markić},\ and\ \citenamefont
  {Boronat}}]{Cikojevi2020}%
  \BibitemOpen
  \bibfield  {author} {\bibinfo {author} {\bibfnamefont {V.}~\bibnamefont
  {Cikojević}}, \bibinfo {author} {\bibfnamefont {L.~V.}\ \bibnamefont
  {Markić}},\ and\ \bibinfo {author} {\bibfnamefont {J.}~\bibnamefont
  {Boronat}},\ }\bibfield  {title} {\bibinfo {title} {Finite-range effects in
  ultradilute quantum drops},\ }\href
  {https://doi.org/10.1088/1367-2630/ab867a} {\bibfield  {journal} {\bibinfo
  {journal} {New J. Phys.}\ }\textbf {\bibinfo {volume} {22}},\ \bibinfo
  {pages} {053045} (\bibinfo {year} {2020})}\BibitemShut {NoStop}%
\bibitem [{\citenamefont {Cikojevi\ifmmode~\acute{c}\else \'{c}\fi{}}\ \emph
  {et~al.}(2020)\citenamefont {Cikojevi\ifmmode~\acute{c}\else \'{c}\fi{}},
  \citenamefont {Marki\ifmmode~\acute{c}\else \'{c}\fi{}}, \citenamefont {Pi},
  \citenamefont {Barranco},\ and\ \citenamefont {Boronat}}]{Cikojevi2020b}%
  \BibitemOpen
  \bibfield  {author} {\bibinfo {author} {\bibfnamefont {V.}~\bibnamefont
  {Cikojevi\ifmmode~\acute{c}\else \'{c}\fi{}}}, \bibinfo {author}
  {\bibfnamefont {L.~V. c.~v.}\ \bibnamefont {Marki\ifmmode~\acute{c}\else
  \'{c}\fi{}}}, \bibinfo {author} {\bibfnamefont {M.}~\bibnamefont {Pi}},
  \bibinfo {author} {\bibfnamefont {M.}~\bibnamefont {Barranco}},\ and\
  \bibinfo {author} {\bibfnamefont {J.}~\bibnamefont {Boronat}},\ }\bibfield
  {title} {\bibinfo {title} {Towards a quantum {M}onte {C}arlo--based density
  functional including finite-range effects: Excitation modes of a
  $^{39}\mathrm{K}$ quantum droplet},\ }\href
  {https://doi.org/10.1103/PhysRevA.102.033335} {\bibfield  {journal} {\bibinfo
   {journal} {Phys. Rev. A}\ }\textbf {\bibinfo {volume} {102}},\ \bibinfo
  {pages} {033335} (\bibinfo {year} {2020})}\BibitemShut {NoStop}%
\bibitem [{\citenamefont {Parisi}\ \emph {et~al.}(2019)\citenamefont {Parisi},
  \citenamefont {Astrakharchik},\ and\ \citenamefont {Giorgini}}]{Parisi2019}%
  \BibitemOpen
  \bibfield  {author} {\bibinfo {author} {\bibfnamefont {L.}~\bibnamefont
  {Parisi}}, \bibinfo {author} {\bibfnamefont {G.~E.}\ \bibnamefont
  {Astrakharchik}},\ and\ \bibinfo {author} {\bibfnamefont {S.}~\bibnamefont
  {Giorgini}},\ }\bibfield  {title} {\bibinfo {title} {Liquid state of
  one-dimensional {B}ose mixtures: A quantum {M}onte {C}arlo study},\ }\href
  {https://doi.org/10.1103/PhysRevLett.122.105302} {\bibfield  {journal}
  {\bibinfo  {journal} {Phys. Rev. Lett.}\ }\textbf {\bibinfo {volume} {122}},\
  \bibinfo {pages} {105302} (\bibinfo {year} {2019})}\BibitemShut {NoStop}%
\bibitem [{\citenamefont {Lorenzi}\ \emph {et~al.}(2023)\citenamefont
  {Lorenzi}, \citenamefont {Bardin},\ and\ \citenamefont
  {Salasnich}}]{Lorenzi2023}%
  \BibitemOpen
  \bibfield  {author} {\bibinfo {author} {\bibfnamefont {F.}~\bibnamefont
  {Lorenzi}}, \bibinfo {author} {\bibfnamefont {A.}~\bibnamefont {Bardin}},\
  and\ \bibinfo {author} {\bibfnamefont {L.}~\bibnamefont {Salasnich}},\
  }\bibfield  {title} {\bibinfo {title} {{On-shell approximation for the
  $s$-wave scattering theory}},\ }\href
  {https://doi.org/10.1103/PhysRevA.107.033325} {\bibfield  {journal} {\bibinfo
   {journal} {Phys. Rev. A}\ }\textbf {\bibinfo {volume} {107}},\ \bibinfo
  {pages} {033325} (\bibinfo {year} {2023})}\BibitemShut {NoStop}%
\bibitem [{\citenamefont {Chiquillo}(2018)}]{Chiquillo2018}%
  \BibitemOpen
  \bibfield  {author} {\bibinfo {author} {\bibfnamefont {E.}~\bibnamefont
  {Chiquillo}},\ }\bibfield  {title} {\bibinfo {title} {{Equation of state of
  the one- and three-dimensional {B}ose-{B}ose gases}},\ }\href
  {https://doi.org/10.1103/PhysRevA.97.063605} {\bibfield  {journal} {\bibinfo
  {journal} {Phys. Rev. A}\ }\textbf {\bibinfo {volume} {97}},\ \bibinfo
  {pages} {063605} (\bibinfo {year} {2018})}\BibitemShut {NoStop}%
\bibitem [{\citenamefont {Ye}\ \emph {et~al.}(2025)\citenamefont {Ye},
  \citenamefont {Zhang}, \citenamefont {Zhou},\ and\ \citenamefont
  {Liang}}]{Ye2026}%
  \BibitemOpen
  \bibfield  {author} {\bibinfo {author} {\bibfnamefont {X.}~\bibnamefont
  {Ye}}, \bibinfo {author} {\bibfnamefont {Y.}~\bibnamefont {Zhang}}, \bibinfo
  {author} {\bibfnamefont {Z.}~\bibnamefont {Zhou}},\ and\ \bibinfo {author}
  {\bibfnamefont {Z.}~\bibnamefont {Liang}},\ }\bibfield  {title} {\bibinfo
  {title} {Tuning spin-density separation via finite-range interactions:
  dimensionality-driven signatures in dynamic structure factors},\ }\href
  {https://doi.org/10.1088/1572-9494/ae0a3d} {\bibfield  {journal} {\bibinfo
  {journal} {Commun. Theor. Phys.}\ }\textbf {\bibinfo {volume} {78}},\
  \bibinfo {pages} {025502} (\bibinfo {year} {2025})}\BibitemShut {NoStop}%
\bibitem [{\citenamefont {Lieb}\ and\ \citenamefont
  {Liniger}(1963)}]{Lieb1963}%
  \BibitemOpen
  \bibfield  {author} {\bibinfo {author} {\bibfnamefont {E.~H.}\ \bibnamefont
  {Lieb}}\ and\ \bibinfo {author} {\bibfnamefont {W.}~\bibnamefont {Liniger}},\
  }\bibfield  {title} {\bibinfo {title} {{Exact analysis of an interacting Bose
  gas. I. The general solution and the ground state}},\ }\href
  {https://doi.org/10.1103/PhysRev.130.1605} {\bibfield  {journal} {\bibinfo
  {journal} {Phys. Rev.}\ }\textbf {\bibinfo {volume} {130}},\ \bibinfo {pages}
  {1605} (\bibinfo {year} {1963})}\BibitemShut {NoStop}%
\bibitem [{\citenamefont {Chin}\ \emph {et~al.}(2010)\citenamefont {Chin},
  \citenamefont {Grimm}, \citenamefont {Julienne},\ and\ \citenamefont
  {Tiesinga}}]{Chin2010}%
  \BibitemOpen
  \bibfield  {author} {\bibinfo {author} {\bibfnamefont {C.}~\bibnamefont
  {Chin}}, \bibinfo {author} {\bibfnamefont {R.}~\bibnamefont {Grimm}},
  \bibinfo {author} {\bibfnamefont {P.}~\bibnamefont {Julienne}},\ and\
  \bibinfo {author} {\bibfnamefont {E.}~\bibnamefont {Tiesinga}},\ }\bibfield
  {title} {\bibinfo {title} {Feshbach resonances in ultracold gases},\ }\href
  {https://doi.org/10.1103/RevModPhys.82.1225} {\bibfield  {journal} {\bibinfo
  {journal} {Rev. Mod. Phys.}\ }\textbf {\bibinfo {volume} {82}},\ \bibinfo
  {pages} {1225} (\bibinfo {year} {2010})}\BibitemShut {NoStop}%
\bibitem [{\citenamefont {Wu}\ and\ \citenamefont
  {Thomas}(2012{\natexlab{a}})}]{Wu2012}%
  \BibitemOpen
  \bibfield  {author} {\bibinfo {author} {\bibfnamefont {H.}~\bibnamefont
  {Wu}}\ and\ \bibinfo {author} {\bibfnamefont {J.~E.}\ \bibnamefont
  {Thomas}},\ }\bibfield  {title} {\bibinfo {title} {{Optical control of the
  scattering length and effective range for magnetically tunable {F}eshbach
  resonances in ultracold gases}},\ }\href
  {https://doi.org/10.1103/PhysRevA.86.063625} {\bibfield  {journal} {\bibinfo
  {journal} {Phys. Rev. A}\ }\textbf {\bibinfo {volume} {86}},\ \bibinfo
  {pages} {063625} (\bibinfo {year} {2012}{\natexlab{a}})}\BibitemShut
  {NoStop}%
\bibitem [{\citenamefont {Wu}\ and\ \citenamefont
  {Thomas}(2012{\natexlab{b}})}]{Haibin2012}%
  \BibitemOpen
  \bibfield  {author} {\bibinfo {author} {\bibfnamefont {H.}~\bibnamefont
  {Wu}}\ and\ \bibinfo {author} {\bibfnamefont {J.~E.}\ \bibnamefont
  {Thomas}},\ }\bibfield  {title} {\bibinfo {title} {Optical control of
  {F}eshbach resonances in {F}ermi gases using molecular dark states},\ }\href
  {https://doi.org/10.1103/PhysRevLett.108.010401} {\bibfield  {journal}
  {\bibinfo  {journal} {Phys. Rev. Lett.}\ }\textbf {\bibinfo {volume} {108}},\
  \bibinfo {pages} {010401} (\bibinfo {year} {2012}{\natexlab{b}})}\BibitemShut
  {NoStop}%
\bibitem [{\citenamefont {Inouye}\ \emph {et~al.}(1998)\citenamefont {Inouye},
  \citenamefont {Andrews}, \citenamefont {Stenger}, \citenamefont {Miesner},
  \citenamefont {Stamper-Kurn},\ and\ \citenamefont {Ketterle}}]{Inouye1998}%
  \BibitemOpen
  \bibfield  {author} {\bibinfo {author} {\bibfnamefont {S.}~\bibnamefont
  {Inouye}}, \bibinfo {author} {\bibfnamefont {M.~R.}\ \bibnamefont {Andrews}},
  \bibinfo {author} {\bibfnamefont {J.}~\bibnamefont {Stenger}}, \bibinfo
  {author} {\bibfnamefont {H.~J.}\ \bibnamefont {Miesner}}, \bibinfo {author}
  {\bibfnamefont {D.~M.}\ \bibnamefont {Stamper-Kurn}},\ and\ \bibinfo {author}
  {\bibfnamefont {W.}~\bibnamefont {Ketterle}},\ }\bibfield  {title} {\bibinfo
  {title} {{Observation of {F}eshbach resonances in a {B}ose--{E}instein
  condensate}},\ }\href {https://doi.org/10.1038/32354} {\bibfield  {journal}
  {\bibinfo  {journal} {Nature (London)}\ }\textbf {\bibinfo {volume} {392}},\
  \bibinfo {pages} {151} (\bibinfo {year} {1998})}\BibitemShut {NoStop}%
\bibitem [{\citenamefont {Tanzi}\ \emph {et~al.}(2018)\citenamefont {Tanzi},
  \citenamefont {Cabrera}, \citenamefont {Sanz}, \citenamefont {Cheiney},
  \citenamefont {Tomza},\ and\ \citenamefont {Tarruell}}]{Tanzi2018}%
  \BibitemOpen
  \bibfield  {author} {\bibinfo {author} {\bibfnamefont {L.}~\bibnamefont
  {Tanzi}}, \bibinfo {author} {\bibfnamefont {C.~R.}\ \bibnamefont {Cabrera}},
  \bibinfo {author} {\bibfnamefont {J.}~\bibnamefont {Sanz}}, \bibinfo {author}
  {\bibfnamefont {P.}~\bibnamefont {Cheiney}}, \bibinfo {author} {\bibfnamefont
  {M.}~\bibnamefont {Tomza}},\ and\ \bibinfo {author} {\bibfnamefont
  {L.}~\bibnamefont {Tarruell}},\ }\bibfield  {title} {\bibinfo {title}
  {{Feshbach resonances in potassium Bose-Bose mixtures}},\ }\href
  {https://doi.org/10.1103/PhysRevA.98.062712} {\bibfield  {journal} {\bibinfo
  {journal} {Phys. Rev. A}\ }\textbf {\bibinfo {volume} {98}},\ \bibinfo
  {pages} {062712} (\bibinfo {year} {2018})}\BibitemShut {NoStop}%
\bibitem [{\citenamefont {Knoop}\ \emph {et~al.}(2011)\citenamefont {Knoop},
  \citenamefont {Schuster}, \citenamefont {Scelle}, \citenamefont {Trautmann},
  \citenamefont {Appmeier}, \citenamefont {Oberthaler}, \citenamefont
  {Tiesinga},\ and\ \citenamefont {Tiemann}}]{Knoop2011}%
  \BibitemOpen
  \bibfield  {author} {\bibinfo {author} {\bibfnamefont {S.}~\bibnamefont
  {Knoop}}, \bibinfo {author} {\bibfnamefont {T.}~\bibnamefont {Schuster}},
  \bibinfo {author} {\bibfnamefont {R.}~\bibnamefont {Scelle}}, \bibinfo
  {author} {\bibfnamefont {A.}~\bibnamefont {Trautmann}}, \bibinfo {author}
  {\bibfnamefont {J.}~\bibnamefont {Appmeier}}, \bibinfo {author}
  {\bibfnamefont {M.~K.}\ \bibnamefont {Oberthaler}}, \bibinfo {author}
  {\bibfnamefont {E.}~\bibnamefont {Tiesinga}},\ and\ \bibinfo {author}
  {\bibfnamefont {E.}~\bibnamefont {Tiemann}},\ }\bibfield  {title} {\bibinfo
  {title} {{Feshbach spectroscopy and analysis of the interaction potentials of
  ultracold sodium}},\ }\href {https://doi.org/10.1103/PhysRevA.83.042704}
  {\bibfield  {journal} {\bibinfo  {journal} {Phys. Rev. A}\ }\textbf {\bibinfo
  {volume} {83}},\ \bibinfo {pages} {042704} (\bibinfo {year}
  {2011})}\BibitemShut {NoStop}%
\bibitem [{\citenamefont {Hu}\ and\ \citenamefont {Liang}(2011)}]{Hu2011}%
  \BibitemOpen
  \bibfield  {author} {\bibinfo {author} {\bibfnamefont {Y.}~\bibnamefont
  {Hu}}\ and\ \bibinfo {author} {\bibfnamefont {Z.}~\bibnamefont {Liang}},\
  }\bibfield  {title} {\bibinfo {title} {Visualization of dimensional effects
  in collective excitations of optically trapped quasi-two-dimensional {B}ose
  gases},\ }\href {https://doi.org/10.1103/PhysRevLett.107.110401} {\bibfield
  {journal} {\bibinfo  {journal} {Phys. Rev. Lett.}\ }\textbf {\bibinfo
  {volume} {107}},\ \bibinfo {pages} {110401} (\bibinfo {year}
  {2011})}\BibitemShut {NoStop}%
\bibitem [{\citenamefont {Hu}\ \emph {et~al.}(2010)\citenamefont {Hu},
  \citenamefont {Liang},\ and\ \citenamefont {Hu}}]{Liang2010}%
  \BibitemOpen
  \bibfield  {author} {\bibinfo {author} {\bibfnamefont {Y.}~\bibnamefont
  {Hu}}, \bibinfo {author} {\bibfnamefont {Z.}~\bibnamefont {Liang}},\ and\
  \bibinfo {author} {\bibfnamefont {B.}~\bibnamefont {Hu}},\ }\bibfield
  {title} {\bibinfo {title} {Collective excitations of a trapped
  {B}ose-{E}instein condensate in the presence of weak disorder and a
  two-dimensional optical lattice},\ }\href
  {https://doi.org/10.1103/PhysRevA.81.053621} {\bibfield  {journal} {\bibinfo
  {journal} {Phys. Rev. A}\ }\textbf {\bibinfo {volume} {81}},\ \bibinfo
  {pages} {053621} (\bibinfo {year} {2010})}\BibitemShut {NoStop}%
\bibitem [{\citenamefont {Wu}\ and\ \citenamefont {Liang}(2018)}]{Wu2018}%
  \BibitemOpen
  \bibfield  {author} {\bibinfo {author} {\bibfnamefont {R.}~\bibnamefont
  {Wu}}\ and\ \bibinfo {author} {\bibfnamefont {Z.}~\bibnamefont {Liang}},\
  }\bibfield  {title} {\bibinfo {title} {Beliaev damping of a
  spin-orbit-coupled {B}ose-{E}instein condensate},\ }\href
  {https://doi.org/10.1103/PhysRevLett.121.180401} {\bibfield  {journal}
  {\bibinfo  {journal} {Phys. Rev. Lett.}\ }\textbf {\bibinfo {volume} {121}},\
  \bibinfo {pages} {180401} (\bibinfo {year} {2018})}\BibitemShut {NoStop}%
\bibitem [{\citenamefont {Stringari}(1996)}]{Stringari1996}%
  \BibitemOpen
  \bibfield  {author} {\bibinfo {author} {\bibfnamefont {S.}~\bibnamefont
  {Stringari}},\ }\bibfield  {title} {\bibinfo {title} {Collective excitations
  of a trapped {B}ose-condensed gas},\ }\href
  {https://doi.org/10.1103/PhysRevLett.77.2360} {\bibfield  {journal} {\bibinfo
   {journal} {Phys. Rev. Lett.}\ }\textbf {\bibinfo {volume} {77}},\ \bibinfo
  {pages} {2360} (\bibinfo {year} {1996})}\BibitemShut {NoStop}%
\bibitem [{\citenamefont {Pitaevskii}\ and\ \citenamefont
  {Stringari}(1998)}]{Pitaevskii1998}%
  \BibitemOpen
  \bibfield  {author} {\bibinfo {author} {\bibfnamefont {L.}~\bibnamefont
  {Pitaevskii}}\ and\ \bibinfo {author} {\bibfnamefont {S.}~\bibnamefont
  {Stringari}},\ }\bibfield  {title} {\bibinfo {title} {Elementary excitations
  in trapped {B}ose-{E}instein condensed gases beyond the mean-field
  approximation},\ }\href {https://doi.org/10.1103/PhysRevLett.81.4541}
  {\bibfield  {journal} {\bibinfo  {journal} {Phys. Rev. Lett.}\ }\textbf
  {\bibinfo {volume} {81}},\ \bibinfo {pages} {4541} (\bibinfo {year}
  {1998})}\BibitemShut {NoStop}%
\bibitem [{\citenamefont {Hu}\ and\ \citenamefont
  {Liu}(2020{\natexlab{c}})}]{Hu2020c}%
  \BibitemOpen
  \bibfield  {author} {\bibinfo {author} {\bibfnamefont {H.}~\bibnamefont
  {Hu}}\ and\ \bibinfo {author} {\bibfnamefont {X.-J.}\ \bibnamefont {Liu}},\
  }\bibfield  {title} {\bibinfo {title} {Collective excitations of a spherical
  ultradilute quantum droplet},\ }\href
  {https://doi.org/10.1103/PhysRevA.102.053303} {\bibfield  {journal} {\bibinfo
   {journal} {Phys. Rev. A}\ }\textbf {\bibinfo {volume} {102}},\ \bibinfo
  {pages} {053303} (\bibinfo {year} {2020}{\natexlab{c}})}\BibitemShut
  {NoStop}%
\bibitem [{\citenamefont {Zhang}\ and\ \citenamefont {Yin}(2023)}]{Zhang2023}%
  \BibitemOpen
  \bibfield  {author} {\bibinfo {author} {\bibfnamefont {F.}~\bibnamefont
  {Zhang}}\ and\ \bibinfo {author} {\bibfnamefont {L.}~\bibnamefont {Yin}},\
  }\bibfield  {title} {\bibinfo {title} {Hydrodynamics of a multi-component
  bosonic superfluid},\ }\href {https://doi.org/10.1088/0256-307X/40/6/066701}
  {\bibfield  {journal} {\bibinfo  {journal} {Chin. Phys. Lett.}\ }\textbf
  {\bibinfo {volume} {40}},\ \bibinfo {pages} {066701} (\bibinfo {year}
  {2023})}\BibitemShut {NoStop}%
\bibitem [{\citenamefont {Mukherjee}\ \emph {et~al.}(2025)\citenamefont
  {Mukherjee}, \citenamefont {Saha},\ and\ \citenamefont
  {Dasgupta}}]{Mukherjee2025}%
  \BibitemOpen
  \bibfield  {author} {\bibinfo {author} {\bibfnamefont {A.}~\bibnamefont
  {Mukherjee}}, \bibinfo {author} {\bibfnamefont {S.}~\bibnamefont {Saha}},\
  and\ \bibinfo {author} {\bibfnamefont {R.}~\bibnamefont {Dasgupta}},\
  }\bibfield  {title} {\bibinfo {title} {Collective excitations in
  two-component ultracold quantum matter},\ }\href
  {https://doi.org/10.1088/1361-648X/addc28} {\bibfield  {journal} {\bibinfo
  {journal} {J. Phys. C}\ }\textbf {\bibinfo {volume} {37}},\ \bibinfo {pages}
  {253003} (\bibinfo {year} {2025})}\BibitemShut {NoStop}%
\bibitem [{\citenamefont {Tylutki}\ \emph {et~al.}(2020)\citenamefont
  {Tylutki}, \citenamefont {Astrakharchik}, \citenamefont {Malomed},\ and\
  \citenamefont {Petrov}}]{Tylutki2020}%
  \BibitemOpen
  \bibfield  {author} {\bibinfo {author} {\bibfnamefont {M.}~\bibnamefont
  {Tylutki}}, \bibinfo {author} {\bibfnamefont {G.~E.}\ \bibnamefont
  {Astrakharchik}}, \bibinfo {author} {\bibfnamefont {B.~A.}\ \bibnamefont
  {Malomed}},\ and\ \bibinfo {author} {\bibfnamefont {D.~S.}\ \bibnamefont
  {Petrov}},\ }\bibfield  {title} {\bibinfo {title} {Collective excitations of
  a one-dimensional quantum droplet},\ }\href
  {https://doi.org/10.1103/PhysRevA.101.051601} {\bibfield  {journal} {\bibinfo
   {journal} {Phys. Rev. A}\ }\textbf {\bibinfo {volume} {101}},\ \bibinfo
  {pages} {051601} (\bibinfo {year} {2020})}\BibitemShut {NoStop}%
\bibitem [{\citenamefont {Andreev}(2025)}]{S.V.2025}%
  \BibitemOpen
  \bibfield  {author} {\bibinfo {author} {\bibfnamefont {S.}~\bibnamefont
  {Andreev}},\ }\bibfield  {title} {\bibinfo {title} {Coupled oscillator model
  of a trapped {F}ermi gas at the {BEC–BCS} crossover},\ }\href
  {https://doi.org/https://doi.org/10.1016/j.physa.2025.130683} {\bibfield
  {journal} {\bibinfo  {journal} {Physica A}\ }\textbf {\bibinfo {volume}
  {674}},\ \bibinfo {pages} {130683} (\bibinfo {year} {2025})}\BibitemShut
  {NoStop}%
\bibitem [{\citenamefont {Hu}\ \emph {et~al.}(2014)\citenamefont {Hu},
  \citenamefont {Xianlong},\ and\ \citenamefont {Liu}}]{Hu2014}%
  \BibitemOpen
  \bibfield  {author} {\bibinfo {author} {\bibfnamefont {H.}~\bibnamefont
  {Hu}}, \bibinfo {author} {\bibfnamefont {G.}~\bibnamefont {Xianlong}},\ and\
  \bibinfo {author} {\bibfnamefont {X.-J.}\ \bibnamefont {Liu}},\ }\bibfield
  {title} {\bibinfo {title} {Collective modes of a one-dimensional trapped
  atomic {B}ose gas at finite temperatures},\ }\href
  {https://doi.org/10.1103/PhysRevA.90.013622} {\bibfield  {journal} {\bibinfo
  {journal} {Phys. Rev. A}\ }\textbf {\bibinfo {volume} {90}},\ \bibinfo
  {pages} {013622} (\bibinfo {year} {2014})}\BibitemShut {NoStop}%
\end{thebibliography}%
\end{document}